\documentclass[twocolumn,prb,showpacs,preprintnumbers,amsmath,amssymb]{revtex4}
\usepackage{amsmath,amssymb,amsfonts,graphicx,bm,units,color, palatino,wrapfig,hyperref,cleveref}
\usepackage{multirow}
\usepackage{bm,bbm,dsfont}       

\newcommand{\bra}[1]{\langle #1 |}
\newcommand{\ket}[1]{| #1 \rangle}

\newcommand{\be}{\begin{equation}}
\newcommand{\ee}{\end{equation}}
\newcommand{\bma}{\begin{pmatrix}}
\newcommand{\ema}{\end{pmatrix}}
\newcommand{\balig}{\begin{align}}
\newcommand{\ealig}{\end{align}}
\newcommand{\ch}{\mathcal{H}}

\newcommand{\bZ}{\mathbb{Z}}
\newcommand{\bI}{\mathbbm{1}}
\newcommand{\dI}{\mathds{1}}
\newcommand{\ba}{\begin{eqnarray}}
\newcommand{\ea}{\end{eqnarray}}

\newcommand{\ignore}[1]{}

\begin{document}

\title{Nontrivial surface topological physics from strong and weak topological insulators and superconductors}
\date{\today}

\author{Ching-Kai Chiu}
\email{chiu7@phas.ubc.ca}
\affiliation{Department of Physics and Astronomy, University of British Columbia, Vancouver, BC, Canada V6T 1Z1} 
\affiliation{Quantum Matter Institute, University of British Columbia, Vancouver BC, Canada V6T 1Z4}

\begin{abstract}
We investigate  states on the surface of strong and weak topological insulators and superconductors that have been gapped by a symmetry breaking term. The surface of a strong 3D topological insulator gapped by a magnetic material is well known to possess a half quantum Hall effect. Furthermore, it has been known that the surface of a weak 3D topological insulator gapped by a charge density wave exhibits a half quantum spin Hall effect. To generalize these results to all Altland-Zirnbauer symmetry classes of topological insulators and superconductors, we reproduce the classification table for the ten symmetry classes by using the representation theory of Clifford algebras and construct minimal-size Dirac Hamiltonians. We find that if the surface dimension and symmetry class possesses a $\mathbb{Z}$ or $\mathbb{Z}_2$ topological invariant, then the resulting surface state with a gapped symmetry breaking term may have a nontrivial topological phase. 

\end{abstract}

\maketitle


\newtheorem{Lemma}{Lemma} \newtheorem{Theorem}[Lemma]{Theorem} \newtheorem{Definition}[Lemma]{Definition} 
\newtheorem{Example}[Lemma]{Example} \newtheorem{Corollary}[Lemma]{Corollary}

\section{Introduction}

	Topological insulators and superconductors (TISC) are fermionic systems with bulk energy gaps separating the occupied and empty bands. They have gapless boundary states that are topologically protected and are related to physical quantities, such as conductivity.  The Hamiltonians of such systems may possess time reversal (TR), particle-hole (PH), or chiral symmetries. The subject started with the recognition, by Thouless {\it et al.\/},\cite{Thouless:1982rz} that the first example of topological insulators is integer quantum Hall states. Twenty years later, Kane and Mele found that by incorporating a spin-orbit coupling in the tight-binding model for graphene\cite{Kane:2005kx, Kane:2005vn}, the system becomes what is now known as a $Z_2$ topological insulator with time reversal symmetry(TRS). Subsequently, Bernevig, Hughes, and Zhang \cite{Bernevig:2006kx} predicted that such $Z_2$ topological insulators can be realized in HgTe/CdTe quantum wells, and this prediction was experimentally verified by K\"{o}nig, {\it et al.\/}\ \cite{MarkusKonig11022007}. After that, 3-D $Z_2$ topological insulators were predicted \cite{Fu:2007fk,Moore:2007uq,Roy:2009kx} and observed  \cite{Chen10072009,Hsieh13022009,Hsieh:2008fk,Xia:2009fk}. It turns out that TR topological insulators are just part of a larger scheme ---  a complete classification of topological insulators and superconductors has been developed by Schnyder {\it et al.\/}\cite{Schnyder:2008gf} and Kitaev \cite{Kitaev}. 
	


In this manuscript we use the representation theory of Clifford algebras\cite{Stone:2011qo,Clifford_reflection} to  construct  minimal-size Dirac Hamiltonians for each symmetry class  and use these Hamiltonians to illustrate boundary-state features. Our results are summarized in \cref{tableminisize} which shows the minimal dimension and number of the inequivalent representations of the Dirac Hamiltonian for each symmetry class and spatial dimension and is in agreement with ref.\ \onlinecite{ribbon_permutation}. We also  write down the explicit matrix form of each of these minimal-model  Dirac Hamiltonians. Such models serve as building blocks for other non-minimal Hamiltonians.  In our  construction, the class  BDI in one dimension and class D in two dimensions, both of which have $\bZ$ topological invariants, are in some sense fundamental. Class BDI in one dimension  serves to  generate all Dirac Hamiltonian possessing  a $\bZ_2$ invariant in odd spacial dimensions. Similarly, class D in two dimensions  generates the non trivial Hamiltonians in even spatial dimensions.

	We use our model Hamiltonians as a tool to investigate states on the surface of both strong and weak TISCs that have been gapped by a magnetic material or by some density wave respectively. The former breaks time reversal symmetry and the latter breaks translational symmetry. The surface of a  3D $\bZ_2$ strong topological insulator  gapped by a magnetic material is well known to possess a half quantum Hall effect. Similarly, a recent paper \cite{Liu:2011fk,Zohar_weak} shows that the surface of a weak 3D topological insulator gapped by a charge density wave (CDW) also has a half-quantum spin Hall effect. We generalize these results to all Altland-Zirnbauer symmetry classes of topological insulators and superconductors in any dimension. When the surface  states are   gapped  by a non-spatial {\it symmetry-breaking\/}  perturbation the surface system will  inevitably belong to a different  symmetry class with fewer symmetries. However, for {\it weak\/}  topological insulators and superconductors which are gapped by non-spatial symmetry-preserving perturbations  (such as a CDW),  the surface system remains  in the {\it same\/}  symmetry class. Significantly, we find that if in the surface dimension and that symmetry class, there is a $\mathbb{Z}_2$ topological invariant, then the resulting surface states must be in a non-trivial phase.
	

	This manuscript is organized as follow. In \cref{classfysymmetry}, we review the ten Zirnbauer-Altland symmetry classes\cite{Zirnbauer:1996fk,PhysRevB.55.1142}. In \cref{complexclass}, we focus on the bulk phase properties of class AIII and A systems by using the representation theory of the complex Clifford algebra to reproduce the first two rows in the classification table (\cref{table:topo}). 
In \cref{real}, similarly, the classification of the remaining eight symmetry classes are reproduced by using the representation theory of real Clifford algebras. In \cref{explicitform}, we provide the rules for constructing the explicit form of each Dirac Hamiltonian. In \cref{3dTI}, we review the examples of surface states for a 3D strong and weak topological insulator, We show that the surface states of a strong topological insulator gapped by ferromagnetic materials have a half Quantum Hall effect and the surface states of a weak topological insulator gapped by CDW have a half Quantum spin Hall effect. In \cref{strongreduction}, we consider more general cases --- the surface states for each strong topological insulators and superconductors gapped by one symmetry breaking. In \cref{weakreduction}, likewise we investigate the topological physics of the surface states for each weak topological insulators and superconductors gapped by some density wave and conclude with a brief summary in \cref{conclusion}. 


\begin{table}
\begin{center}
\begin{tabular}{ |c | | c | c | c || c| c|c|c|c|}
\hline			
 Class &   $TR$ &  $PH$& $Ch$  &   d=0 & d=1 & d=2 & d=3 & d=4 \\
\hline 
\hline
AIII &   0 & 0 &1 	 			&   0 &$\mathbb{Z}$ & 0 & $\mathbb{Z}$ & 0\\
\hline
A &   0 & 0 &0   								&   $\mathbb{Z}$  & 0 &$\mathbb{Z}$ & 0 & $\mathbb{Z}$ \\
\hline	
\hline		
  D &   0 & +1 & 0   	 						&  $\mathbb{Z}_2$ & $\mathbb{Z}_2$ & $\mathbb{Z}$ &0 & 0 \\
\hline  
  DIII &    -1 & +1 & 1  					&   0  & $\mathbb{Z}_2$ & $\mathbb{Z}_2$ & $\mathbb{Z}$ & 0 \\
\hline  
  AII &   -1 &  0  	 					& 0 &    $2\mathbb{Z}$ & 0 & $\mathbb{Z}_2$  & $\mathbb{Z}_2$ & $\mathbb{Z}$\\
\hline  
  CII &   -1 & -1 & 1 	  		&  0 & $2\mathbb{Z}$ &  $0$ & $\mathbb{Z}_2$ & $\mathbb{Z}_2$  \\
\hline  
  C &  0  & -1 & 0 	 						&   0 & 0  &$2\mathbb{Z}$ & 0 & $\mathbb{Z}_2$ \\
\hline
  CI &   +1  & -1 & 1  					&   $0$ & 0 &0  & $2\mathbb{Z}$ & 0\\
\hline 
   AI &   +1 & 0  & 0  					&   $\mathbb{Z}$ & 0& $0$ &0 & $2\mathbb{Z}$ \\
\hline
   BDI &    +1  & +1 & 1   		&   $\mathbb{Z}_2$ & $\mathbb{Z}$ & 0 & 0 & 0 \\
\hline
\end{tabular}
\end{center}
\caption{\cite{Kitaev,Schnyder:2008gf} The first column represents the names of the ten symmetry classes, associated with the presence or absence of TR, PH, and chiral symmetries in the next three columns. The number $0$ in the next three columns denotes the absence of the symmetry. The numbers $+1$ and $-1$ denote the presence of the symmetry and indicate the signs of the square TR operator and the square PH operator.  }
\label{table:topo}
\end{table}

\section{Classification by Symmetries}\label{classfysymmetry}

 	We start from single-particle Hamiltonian $\ch(\vec{k})$ in momentum space and adopt the notations in ref. \onlinecite{chiu_reflection_gapped,chiu_reflection_gapless}.
 
 	The Dyson symmetry classes, completed by Zirnbauer and Altland, are simply characterized by the presence or absence of three discrete symmetries: time reversal, particle-hole, and chiral (sublattice) symmetries. The time reversal operator $T$ is an antilinear map with $T^2=\pm 1$. If the system preserves TRS, the Hamiltonian $H$ obeys the equation 
\begin{equation}
T^{-1}\ch(-\vec{k})T=\ch(\vec{k}), \label{time}
\end{equation}  
where $\vec{k}$ is the momentum. Similarly, $C$ is a particle-hole operator as an antilinear map with $C^2=\pm 1$. The PHS equation can be written in the similar form
\begin{equation}
C^{-1}\ch(\vec{k})C=-\ch(\vec{k}). \label{PH}
\end{equation}
Finally, a Hamiltonian preserving chiral symmetry satisfies the equation
\begin{equation}
S^{-1}\ch(\vec{k})S=-\ch(\vec{k}), \label{sublattice}
\end{equation}
where the chiral symmetry operator $S$ is a {\it linear\/} map. TRS and PHS guarantee chiral symmetry with the symmetry operator $S=TC$. However, a Hamiltonian preserving chiral symmetry may not preserve the other two symmetries --- as seen in the class AIII in \cref{table:topo}. Class AIII and Class A (having no symmetry associated with it) are known as \emph{complex} symmetry classes. In \cref{table:topo}, the remaining symmetry classes contain all combinations with either TRS or PHS being preserved. There are two possible values that can be assigned to each symmetry when it is preserved, \emph{i.e.} 
\be
T^2=\pm \dI \text{ and } C^2=\pm \dI. \label{T2C2}
\ee  
These eight symmetry classes are called \emph{real} for the reasons will be clear later.

All ten symmetry classes of electronic systems in \cref{table:topo} are bulk gapped systems. To topologically distinguish any two bulk gapped systems in the same symmetry class, we have to consider a continuous deformation between these two systems. When the spectra of the two systems can be continuously deformed from one to the other without closing the band gap, these two systems possess the same topological phases. To capture the essential topological phases simply, we simplify our focus to Dirac Hamiltonians, which are topologically equivalent to another systems with more complicated band structures. For the Dirac Hamiltonian, we will also require that if the eigenenergies of the conduction band levels are $\{+E\}$, then the eigenenergies of the valence band levels are $\{-E\}$. A Dirac Hamiltonian is a linear combination of Gamma matrices ($\gamma_i$): 
\begin{equation}
\ch_{\mathrm{Dirac}}=M\gamma_{0}+\sum_{i=1}^d k_d \gamma_d,  \label{Hform} 
\end{equation}
where $\gamma_i$'s obey the anticommutation relation of the Clifford algebra
\be
\{\gamma_i,\gamma_j\}=2\delta_{ij}\dI. \label{anticommutation}
\ee
Here, $d$ is the spatial dimensions of the system, $k_i$ is the momentum in the i-th direction, and $\dI$ is the identity matrix without specific matrix dimension. The eigenenergies are given by $E_{\pm}=\pm\sqrt{M^2+\sum_{i=1}^d k_i^2}$. For a system in each individual symmetry class, the Dirac Hamiltonian must preserve the symmetries corresponding to that symmetry class. That is, the Hamiltonian obeys the symmetry equations in \cref{time,PH,sublattice}. The first goal of this manuscript is to find the {\it minimum matrix dimension\/}(size) of a Dirac Hamiltonian in each individual symmetry class and spatial dimension. 

	With negative energies filled, we identify $M=0$ as a quantum phase transition, where the gap between the empty and occupied band closes at $k_i=0$. In general we might worry that, for realistic systems, additional perturbations and disorder, which preserve system's symmetries, might block the quantum phase transition. Such terms must anticommute with all of the gamma matrices in the Dirac Hamiltonian. We classify these terms as an {\it extra mass term\/}, $m\gamma_{d+1}$. The reason is that the new energies are $E_{\pm}=\pm\sqrt{M^2+m^2+\sum_{i=1}^d k_i^2}$. If allowed by symmetries, such $m$ will not vanish in a real system and quantum phase transition cannot take place by tuning $M$. Furthermore, we name this term as symmetry preserving extra mass term (SPEMT). If $m\gamma_{d+1}$ is not allowed by symmetries, the systems with negative $M$ and positive $M$ respectively have two different topological phases.

\subsection{The spectrum of the surface states}\label{surfacecal}
We will show that the Dirac Hamiltonians $H_{\mathrm{Dirac}}$ above have surface states localized on any domain wall where $M$ changes the sign. Furthermore, without an extra mass term these surface states are gapless and with an extra mass term they are gapped. To see those properties, suppose that there is a domain wall at $x_d=0$ (in $d$-th direction). That is, we change $M(x_d)$ from negative to positive as $x_d$ increases. As $k_{d}$ is no longer a good quantum number, we replace $k_{d}$ by $-i\partial/\partial x_d$. First, to have the zero mode $\Phi$, we set $k_j$ to be zero for $j\neq d$, so $H_{\mathrm{Dirac}}\Phi$ becomes
\be
0=(M\gamma_0 -i\gamma_d\partial/\partial x_d)\Phi=\gamma_0(M-i\gamma_0\gamma_d\partial/\partial x_d)\Phi
\ee
The normalizable zero mode is proportional to $e^{-\int ^{x_d}_0M(x_d')dx_d'}$. Furthermore, because $(i\gamma_0\gamma_d)^2=\bI$, $i\gamma_0\gamma_d$ only has $\pm 1$ eigenvalues. Since $i\gamma_0\gamma_d$ and $\gamma_0$ anticommute with each other, the eigenspaces corresponding to the eigenvalues $\pm 1$ have the same dimension. By choosing an eigenvector $\psi_-$ of the eigenvalue $-1$, the normalized zero mode is $\Phi_-=e^{-\int ^{x_d}_0M(x_d')dx_d'}\psi_{-}$. Because $[i\gamma_0\gamma_1,\gamma_j]=0$ for $j\neq d$. The bulk Hamiltonian $H_{\mathrm{Dirac}}$ can be projected onto a surface Hamiltonian
\be
\ch^{\mathrm{surf}}_{\mathrm{Dirac}}=\sum_{i=1}^{d-1}k_i\Upsilon_{i-}, \label{surfH1}
\ee
where the subscript ``$-$'' indicates that $\Upsilon_{i-}$ is in the basis of the eigenspace of $i\gamma_0\gamma_d=-1$; $\Upsilon_{i-}$ is half the size of $\gamma_i$ and still preserves the anticommutation relations in \cref{anticommutation}. Hence, the energy spectrum of the surface Hamiltonian is gapless 
\be
E_{\pm}=\pm\sqrt{\sum_{i=1}^{d-1}k_i^2}.
\ee
However, when an extra mass term, $m\gamma_{d+1}$, is added to $\ch_{\mathrm{Dirac}}$, the surface Hamiltonian is gapped out since another anticommuting term $m\Upsilon_{d+1-}$ is in $\ch^{\mathrm{surf}}$. Returning to the quantum phase transition discussion. The allowance of an extra mass term into $H_{\mathrm{Dirac}}$ implies that the regions of the system $x_d>0$ and $x_d<0$ have the same topological phase because the extra mass term prevents gap closing. We know the fact that the interface between two systems in the same phase have no robust gapless states. This is consistent with the absence of gapless states around $x_d=0$. In the next  two sections, we will discuss topological invariants for complex and real symmetry classes by using the occurrence of quantum phase transitions in bulk Dirac Hamiltonians. 

\section{Complex symmetry classes}\label{complexclass}
The existence of quantum phase transitions can determine the classification of the topological phases for each symmetry classes. The topological phases are classified by three types of the topological invariants: $``0",\ \mathbb{Z}_2,$ and $ \mathbb{Z}$. 
The system that cannot have any quantum phase transitions is a trivial $``0"$ insulator. 
When quantum phase transitions are possible, we need to consider additionally a system with two identical copies of this Hamiltonian. In this new system, if quantum phase transitions cannot take place, it is a $\mathbb{Z}_2$ topological insulator. Conversely, if quantum phase transitions can occur, the system is a $\mathbb{Z}$ topological insulator. These occurrences of quantum phase transition will be analyzed by whether or not SPEMTs are allowed in the Hamiltonian. Before discussing the more complicated cases of symmetry classes with TRS or PHS, let us focus on two simple symmetry classes AIII and A, which preserves and breaks chiral symmetry respectively. They can be directly classified by irreducible matrix representations of the complex Clifford algebra. Class A and AIII are closely related; their topological invariants $\{0,\mathbb{Z}\}$ flip as the spatial dimensions are shifted by one, as shown in \cref{table:topo}.

First, we start with some simple examples. Consider class A, the one without any symmetry, in 1-dimension. The Hamiltonian with gaps in the irreducible ({\it minimum matrix dimension\/}) form of \cref{Hform} is given by
\begin{equation}
\ch=M\sigma_x+p_1\sigma_y \label{1dnosy},
\end{equation}
where M is a parameter to be tuned close the bulk energy gap and $p_1$ is the momentum. In order to switch the phase, the only possible place where the gap closes is at $M=p_1=0$. Because we do not require any symmetry in this system, it is legal to add an extra mass term $m\sigma_z$ $(m\neq 0)$, as a small perturbation, to keep the gap open. Hence, no quantum phase transitions take place. The system is characterized as a single trivial phase insulator. 

For class AIII in 1-D, the system satisfies the chiral symmetry condition in \cref{sublattice}; that is, $S$ anticommutes with the Hamiltonian. The Hamiltonian is in the same form as \cref{1dnosy}, but with $S=\sigma_z$. Now we cannot add $m\sigma_z$ in the Hamiltonian because it commutes with $S$. At $p_1=M=0$ the gap closes safely so $M>0$ and $M<0$ are two different quantum phases. 

To increase the chances of finding an extra mass term, consider two copies of this Hamiltonian in the system, which preserves chiral symmetry. For example, electrons with spin up and down are described by two identical Hamiltonians. The form of the new Hamiltonian is given by
\begin{equation}
\ch=M\sigma_x\otimes \bI+p_1 \sigma_y\otimes \bI, 
\end{equation}
with chiral symmetry operator $S=\sigma_z\otimes\bI$, where $\bI$ is the $2\times 2$ identity matrix. It is not hard to prove that no SPEMTs exist. At $M=p_1=0$, four-band touching brings a different quantum phase. Alternatively, as we change the sign of one term of one $2\times 2$ single electron Hamiltonian in \cref{1dnosy}, the entire Hamiltonian can be rewritten as
\begin{equation}
\ch=M\sigma_x\otimes \sigma_z + p_1 \sigma_y\otimes \bI
\end{equation}
It is not difficult to find an SPEMT $m\sigma_x\otimes \sigma_x$ as a perturbation preserving chiral symmetry. This term prevents the spectrum from closing the gaps by tuning $M$ so the system always stays in the same phase. In general, there are $k+l$ copies of the Hamiltonian in \cref{1dnosy} and $l$ copies of the same Hamiltonian with the first term sign changing and $k$ copies of the original Hamiltonian. Say $k\geq l$. The composite Hamiltonian is expressed as
\begin{equation}
\ch=M\sigma_x\otimes 
\left( \begin{array}{cc}
\sigma_z\otimes \dI_{l\times l} & 0\\
0 & \dI_{(k-l)\times (k-l)}   
\end{array} \right)
+p_1\sigma_y\otimes\dI_{(k+l)\times (k+l)}.
\end{equation}
The symmetry operator $S$ should be enlarged to $\sigma_z\otimes \dI_{(k+l)\times (k+l)}$. The energy gap closing in the first block of the Hamiltonian is forbidden by a SPEMT $(m\sigma_x \otimes \sigma_x \otimes \dI_{k\times k})$. However, by tuning $M$ the energy gap in the second block can close freely to transit to different phases. Therefore, the integer $k-l$ characterizes $k-l$ different topological phases so the characterization provides the integer number of different topological phases labelled by $\mathbb{Z}$ in class AIII in one spatial dimension.

\begin{table}  
\begin{center}
\begin{tabular} {|c | c | c | c| c |c| c|c|c|}
\hline
$q$ & 1 & 2 & 3 & 4 & 5 & 6 & 7 & 8\\
\hline
\hline
$d_q$ & $1_2$ & 2 & $2_2$ & 4 & $4_2$ & 8 & $8_2$ & 16 \\
\hline
\end{tabular}
\end{center}
\caption{The integer $q$ is the number of the complex gamma matrices, which anticommute with the rest of the gamma matrices. The second row indicates the corresponding minimum matrix dimensions of the gamma matrices. The subscript 2 in the second row denotes that there are two inequivalent representations. (\emph{i.e.}\ these two representations cannot unitary transform from one to another. For example,  for $q=3$, $\{\sigma_x,\sigma_y,\sigma_z\}$ and $\{\sigma_x,\sigma_y,-\sigma_z\}$ are two inequivalent representations.) With $q=2n-1$ the sets ($\{\gamma_1,\gamma_2,...,\gamma_{2n-2},\gamma_{2n-1}\}$ and $\{\gamma_1,\gamma_2,...,\gamma_{2n-2},-\gamma_{2n-1}\}$) of the two inequivalent representations allow us to construct a new set of the gamma matrices ($\Gamma_i$) for $q=2n$: $\Gamma_i=\gamma_i\otimes\bI $ for $i\leq 2n-2$, $\Gamma_{2n-1}=\gamma_{2n-1}\otimes\sigma_z $, and $\Gamma_{2n}=\gamma_{2n-1}\otimes\sigma_x $. The new gamma ($\Gamma_{2n}$) matrix anticommutes with the other gamma matrices.}
\label{complexrep}
\end{table}

	In general, for class A in d spatial dimensions, the construction of the Hamiltonian needs $d+1$ gamma matrices. Also for class AIII in d spatial dimensions, the formation needs $d+2$ gamma matrices due to the presence of the chiral symmetry operator. From the representation theory of the complex Clifford algebra, \cref{complexrep} shows the {\it minimum matrix dimension\/}(size) $d_q$ of the Hamiltonian for the number $q$ of irreducible complex gamma matrices is given by 
\begin{equation}
d_q=2^{\lfloor q/2\rfloor  } \label{complexsize},
\end{equation}
where $\lfloor x \rfloor $ equals to the largest integer less than $x$. Therefore, in $d$ spatial dimensions the minimal matrix dimensions of the Hamiltonians for class A and class are given by $2^{\lfloor d/2+1/2\rfloor }$ and $2^{\lfloor d/2+1\rfloor  }$ respectively. We name a Dirac Hamiltonian written in minimal matrix dimensions as a \emph{minimal Dirac Hamiltonian}.  

On the one hand, consider $q$ is even, say $q=2n$, then $d_q=2^n$. 
If we add an extra mass term as a gamma matrix to keep the energy gap open, the matrix dimension $(d_q=2^n)$ of the Hamiltonian is unchanged. Such an extra mass term can be present in the Hamiltonians in class A and AIII because class A corresponds to no symmetries and in class AIII this mass term, an extra gamma matrix, anticommutes with the $S$ chiral symmetry operator. For both of the symmetry classes, due to the presence of the SPEMT by tuning $M$ the systems always stay in the same phase. 
In this regard, like class A in one spatial dimension in \cref{1dnosy} is always in the only one topological phase, which is trivial. 
In general, in class AIII in even dimensions (class A in odd dimensions), a Dirac Hamiltonian with(out) a chiral symmetry operator is constructed by even number of gamma matrices so the systems are characterized by $``0"$ topological invariant.

On the other hand, $q=2n-1$ is odd. It is not possible that an extra gamma matrix is added to a minimal Dirac Hamiltonian without enlarging matrix dimension. The minimal Hamiltonian is safe to pass through phase transition under the symmetry by tuning $M$. To add an SPEMT, \cref{complexrep} displays that the size of the Hamiltonian must be doubled and the new double size gamma matrices must be constructed by the two inequivalent representations of the original gamma matrices $\gamma_i$. Like class AIII in one spatial dimension, the Hamiltonian can be arbitrarily enlarged but is formed by original $\gamma_i$ of the Hamiltonian in irreducible form. 
\begin{equation}
\ch=M\gamma_{2n-2}\otimes 
\left( \begin{array}{cc}
\sigma_z\otimes \mathbb{I}_{k\times k} &   \\
& \mathbb{I}_{l\times l}   
\end{array} \right)
+\sum_i p_i\gamma_{i}\otimes\mathbb{I}_{(2k+l)\times (2k+l)}
\end{equation}
The first block of the mass matrix of the second matrix denotes $k$ pairs of the two inequivalent representations, which allow to anticommute with an SPEMT. Also the second block denotes the $l$ equivalent representations, which provide $l$ times of the quantum phase transitions. Since $l$ can be an arbitrary integer, for class A in even-dimension and class AIII in odd-dimension, the system has a $\mathbb{Z}$ topological invariant.

{ 
\section{Real symmetry classes} \label{real}
The {\it minimum dimension\/} of Dirac Hamiltonians in the presence and absence of an SPEMT for each individual {\it real\/} symmetry class plays an essential role to connect topological invariants: $``0"$, $\bZ$, and $\bZ_2$. In the following, we use Clifford algebra to obtain the \emph{minimum dimension} of Dirac Hamiltonian in each real symmetry class and then compare the \emph{minimum dimension} with and without an SPEMT to determine the topology. Since the SPEMT is important in this topological realization, we slightly change the Dirac Hamiltonian in \cref{Hform} by adding some SPEMTs and change $\gamma_0\rightarrow \tilde{\gamma}_0$
\begin{equation}
\ch({\bf k})=M\tilde{\gamma}_0+\sum_{j=1}^Dm_j\tilde{\gamma}_j+\sum_{i=1}^dk_i\gamma_i, \label{Hmk}
\end{equation}
where $D$ is equal to $0$ or $1$ corresponding to a system without or with an SPEMT and $d$ indicates the spatial dimension of the system. Also these gamma matrices satisfy the anticommutation relations:
\be
\{\gamma_i,\gamma_j\}=2\delta_{ij}\dI,\ \{\tilde{\gamma}_i,\tilde{\gamma}_j\}=2\delta_{ij}\dI,\ \{\gamma_i,\tilde{\gamma}_j\}=0 \label{gammaanti}
\ee
For each individual symmetry class, this Hamiltonian preserves the corresponding symmetries: TRS, PHS, or chiral symmetry. To construct this Hamiltonian, we need to have these elements: 
\be
\mathcal{S}=\{i,\ \tilde{\gamma}_0,\ \tilde{\gamma}_1,\cdots,\ \tilde{\gamma}_D,\ \gamma_1,\ \gamma_2,\cdots,\ \gamma_d,\ T,\ \text{and(or)}\ C\} \label{elements}
\ee
The reason for $i\in \mathcal{S}$ is that $i^2=-1$ guarantees complex structure so the Hamiltonian can be constructed in a complex vector space. Here presence of time reversal operator $T$ and particle-hole operator $C$ depends on which symmetry class the Hamiltonian belongs to. Because of antilinearity, these operators anticommute with $i$
\be
\{i,T\}=0,\ \{i,C\}=0. \label{TCi}
\ee
When the Hamiltonian preserves time reversal symmetry and particle hole , the gamma matrices must obey
\begin{align}
[\tilde{\gamma}_j,T]=0,\ \{\gamma_i,T\}=0 \label{Tc} \\
\{\tilde{\gamma}_j,T\}=0,\ [\gamma_i,C]=0 \label{Cc}
\end{align} 
For real symmetry classes, TRS and PHS automatically guarantee chiral symmetry so no extra elements are required for the construction of chiral operator $S$. 
The operators $T$ and $C$ describe different degree freedoms of physical systems. Due to the antilinear property of these operators, we can add a proper phase to $T$ or $C$ such that 
\be
\{T,C\}=0 \label{TCa}
\ee 
The goal of this section is to find the {\it minimum matrix dimension\/} of the elements in $\mathcal{S}$, which satisfy the anticommutation and commutation relations in \cref{T2C2,gammaanti,TCi,Tc,Cc,TCa}. To achieve this, from $\mathcal{S}$ we construct a group $G$ respecting the anticommutation and commutation relations 
\be
G\equiv\{s_1^{n_1}s_2^{n_2}\cdots s_r^{n_r}|s_i\in\mathcal{S},n_i\in``0"\cup\bZ^+\} \label{groupG}
\ee
In other words, the elements of $\mathcal{S}$ are the generators of $G$. 
According to representation theory, if for a finite group a faithful representation, which is isomorphic to the group, is irreducible, the matrix dimension of the faithful representation is minimal. Therefore, we have to find the irreducible faithful representation of $G$ in order to obtain the {\it minimum dimension\/} of the Dirac Hamiltonian. For the future convenience, we define the $G_{\#}$ from \cref{groupG} corresponding to that symmetry class, where $\#$ is the name of symmetry class. The main idea of this section is the application of the isomorphisms between $G_{\#}$ and the group from the generators of the real Clifford algebra $Cl_{p,q}$. That is, 
\begin{align}
G_{\text{D}}\cong &Cl^g_{2+D,1+d}, & G_{\text{DIII}}\cong &  Cl^g_{3+D,1+d}, \nonumber \\
G_{\text{AII}} \cong &Cl^g_{3+D,d},\ & G_{\text{CII}}\cong &Cl^g_{4+D,d}, \nonumber \\
G_{\text{C}}\cong & Cl^g_{2+d,1+D},\ & G_{\text{CI}}\cong & Cl^g_{2+d,2+D},  \nonumber \\ 
G_{\text{AI}}\cong & Cl^g_{1+d,2+D}, & G_{\text{BDI}}\cong &  Cl^g_{1+d,3+D}  \label{iso}
\end{align}
The group $Cl^g_{p,q>0}$ (not $\mathrm{pin}(p,q)$) is define as
\begin{align}
Cl^g_{p,q}\equiv\{L_1^{n_1}L_2^{n_2}\cdots L_r^{n_r}|L_i\in\mathcal{J},n_i\in``0"\cup\bZ^+\},  
\end{align}
where the set $\mathcal{J}$
\be
\mathcal{J}\equiv\{\tilde{J}_1,\tilde{J}_2,\cdots,\tilde{J}_p,J_1,J_2,\cdots,J_q\}
\ee 
and these generators of $J_i$ and $\tilde{J}_j$ are real symmetric matrices and obey the anticommutation relations. 
\be
\{J_i,J_j\}=-2\delta_{ij}\dI,\ \{\tilde{J}_i,\tilde{J}_j\}=2\delta_{ij}\dI,\ \{J_i,\tilde{J}_j\}=0 \label{Cliffordrelation}
\ee
We note that the Dirac Hamiltonians are in a complex vector space and those generators are in a real vector space. This is the reason that these eight symmetry classes are called {\it real\/}. We leave the proof of this isomorphism in \cref{isomorphism} for the interested readers.

In fact, we are interested in the {\it minimum dimension\/} of a Dirac Hamiltonian for each specific symmetry and spatial dimension. In \cref{isomorphism}, we will show that the {\it minimum dimension\/}  of that Dirac Hamiltonian is a half {\it minimum dimension\/}  $d_{p,q}$ of the corresponding real Clifford algebra in the matrix form due to the transformation from a real vector space to a complex vector space. The minimum dimension of a Dirac Hamiltonian for each symmetry class and spatial dimension is shown in \cref{tableminisize}. The table will be generated as follows. 

Although the isomorphism in \cref{iso} provide a {\it minimum dimension\/} of the Dirac Hamiltonians for each symmetry class and each spatial dimension, a better way to reveal the {\it minimum dimensions\/} is to use some identities of the real Clifford algebra. The first two identities are given by
\begin{align}
Cl_{p+1,q+1}&=Cl_{p,q}\otimes Cl_{1,1} \label{oneplusone} \\
Cl_{p+k,q}&=Cl_{p,q+k},\ \text{as}\ p-q\equiv 3\ (\text{mod}\ 4).
\end{align}
We note that $Cl_{1,1}$ is isomorphic to $R(2)$, where $R(n)$ denotes the algebra of $n\times n$ matrices with coefficients in $\mathbb{R}$. Therefore, the  {\it minimum dimension\/}  $d_{1,1}$ of the generators of the $Cl_{1,1}$ is 2. By the first identity, the {\it minimum dimensions\/}  $d_{p+1,q+1}$ of the generators of $Cl_{p+1,q+1}$ is the product of $d_{1,1}$ and $d_{p,q}$. Thus, these identities above provide these relations of the {\it minimum dimensions\/}: 
\begin{align}
d_{p+1,q+1}&=2d_{p,q} \\
d_{p+k,q}&=d_{p,q+k},\ \text{as}\ p-q\equiv 1\ (\text{mod}\ 4). 
\end{align} 
By using \cref{iso}, we have the the relations for {\it minimum dimensions\/} of the Dirac Hamiltonians in \cref{Hmk} in different symmetry classes as the number $D$ of $\tilde{\gamma}_i$ is fixed 
\begin{align}
&size(\ch_{\rm{D}}(\Delta d=0))=size(\ch_{\rm{DIII}}(\Delta d=1))2^{-1} \nonumber \\
=&size(\ch_{\rm{AII}}(\Delta d=2))2^{-1}=size(\ch_{\rm{CII}}(\Delta d=3))2^{-2} \nonumber \\
=&size(\ch_{\rm{C}}(\Delta d=4))2^{-2}=size(\ch_{\rm{CI}}(\Delta d=5))2^{-3}  \nonumber \\
=&size(\ch_{\rm{AI}}(\Delta d=6))2^{-3}=size(\ch_{\rm{BDI}}(\Delta d=7))2^{-4}, \label{dmove}
\end{align}
where $\Delta d=d-n$ and $n$ is an arbitrary integer. This equation shows that the minimum dimensions of two different symmetry classes can be connected by spatial dimensional $d$ shift. Therefore, since the behaviors of the minimum dimensions determine topology, in the ten-fold classification table (\cref{table:topo}) the topology of a symmetry class can obtained from another symmetry class by dimensional shift. Last but not least, we introduce another identity of the real Clifford algebra: the Bott periodicity 
\be
Cl_{p,q}\otimes Cl_{4,4}=Cl_{p,q+8}=Cl_{p+8,q},
\ee
which provides that the  {\it minimum dimension\/} of a Dirac Hamiltonian for each individual symmetry possesses the periodicity of eight 
\be
size(\ch_\#(\Delta d=0))=size(\ch_\#(\Delta d=8))2^{-4}, \label{eight}
\ee
where $\#$ is the name of a symmetry class. From these relations as D is fixed, by knowing the {\it minimum dimension\/} of a Dirac Hamiltonian as $d=1$ for each symmetry class we know the information for the {\it minimum dimension\/} of a Dirac Hamiltonian in any spatial dimensions and symmetry class. For example, 
\begin{align}
size(\ch_{D}(d=4))=&size(\ch_{CI}(d=9))/8 \nonumber \\ 
=&2\times size(\ch_{CI}(d=1)) \nonumber 
\end{align}
In the first line and the second line, we use \cref{dmove,eight} respectively. The topological invariants $\{0,\bZ_2,\bZ\}$ are determined by the {\it minimum dimensions\/} of $\ch_\#$ with $D=0,\ 1$ as the spatial dimension $d$ is fixed. It will be shown in the next three paragraphs that the changing of the {\it minimum dimensions\/} from $D=0$ to $D=1$, $(n\rightarrow n, n\rightarrow n_2),\ n\rightarrow 2n,$ and $n_2\rightarrow 2n$ correspond to topological invariants $``0",\ \bZ_2,$ and $\bZ$ respectively. This corresponding and the relations of the {\it minimum dimensions\/} in \cref{dmove,eight} show that in the periodic table when spatial dimension $d$ is increased by one, the entire topological invariant column moves down by one. Thus, all we need to know is the topological invariants as $d=1$ for all symmetry classes to generate the periodic table in any spatial dimension. According to the isomorphism in \cref{iso}, the {\it minimum dimension\/} of a Dirac Hamiltonian in one spatial dimension without ($D=0$) and with ($D=1$) an SPEMT in a complex vector space is shown in \cref{size1dc}. Finally, we can regenerate the periodic table in \cref{table:topo} from the last column in \cref{size1dc}, which provides the corresponding topological invariants for each symmetry class.

\begin{table} 
\begin{center} 
\begin{tabular}{ |c || c | c |c|}
\hline			
 Class & $d=1,\ D=0$  &  $d=1,\ D=1$  &   invariant \\
\hline 
\hline		
  D &   2 & 4 & $\bZ_2$\\
\hline  
  DIII &    4 & 4 &$0$ \\
\hline  
  AII &     $2_2$ & 4 &$\bZ$ \\
\hline  
  CII &   4 & 4 & $0$ \\
\hline  
  C &  2 & 2 & $0$ \\
\hline
  CI &  2 & $2_2$ & $0$ \\
\hline 
   AI &   $1_2$ & 2 & $\bZ$\\
\hline
   BDI &   2 & 4 & $\bZ_2$ \\
\hline 
\end{tabular} 
\end{center}
\caption{The second and third column represent the {\it minimum dimension\/} of a Dirac Hamiltonian without and with an SPEMT.} 
\label{size1dc}
\end{table}

	$n\rightarrow n, n\rightarrow n_2$: An SPEMT can be added into the Hamiltonian without enlarging the matrix dimension. Therefore, in the presence of the SPEMT the bulk Dirac Hamiltonian in \cref{Hmk} can go through a quantum phase transition by tuning $M$. 
	 This is a topological trivial case and is labeled by $``0"$. 
	
	$n\rightarrow 2n$:  For the Hamiltonian with the minimum size, an extra mass term is forbidden unless breaking the symmetries. Hence, the gapless state is topologically protected. However, we consider two copies of the bulk Dirac Hamiltonian. As the size of the Hamiltonian is doubled, in the presence of an SPEMT the Hamiltonian is gapped. On one hand, for the odd number of the Dirac Hamiltonian copies in \cref{Hmk}, one of the copies cannot be paired with another to have SPEMTs. Therefore, there is one quantum phase transition, which implies two distinct topological phases. On the other hand, for the even number of the copies, each state can be paired to have SPEMTs, which prevent the occurrence of quantum phase transitions. The absence of quantum phase transitions keeps the system in the only topological phase. Thus, this case belongs to a $\mathbb{Z}_2$ topological invariant. 
	
	$n_2\rightarrow 2n$: A single bulk Dirac Hamiltonian in minimum matrix dimension has two topological phases in the absence of SPEMT. However, when we double the size of the Hamiltonian, there are two possibilities: the new Hamiltonian can be constructed by two equivalent representations or two inequivalent representations from the original one. For the former, SPEMTs do not exist. These two Dirac Hamiltonians, with two equivalent representations, transit to different topological phases. For the latter, SPEMTs is present in the Hamiltonian. Hence, the system of the two Dirac Hamiltonians, with two inequivalent representations, is always in the same topological phase, which is trivial. In general, the Hamiltonian formed by $n$ equivalent representations can go through $n$ times of quantum phase transitions. In this regard, this case gives a $\mathbb{Z}$ topological invariant.

	The topological features of the invariants $\{0,\bZ_2,\bZ\}$ has been also discussed in ref \onlinecite{chiu_reflection_gapped,chiu_reflection_gapless}.

\begin{table} 
\begin{center}
\begin{tabular}{ |c | | c | c | c || c| c|c|c|c|} 
\hline 
& & & &  \multicolumn{5}{c|}{d} \\
\cline{5-9}			
 Class &   $TR$ &  $PH$& $Ch$  &   d=0 & d=1 & d=2 & d=3 & d=4 \\
\hline 
\hline
AIII &   0 & 0 &1 	 			&   2 &$\bf 2_2$ & 4 & $\bf 2^2_2$ & 8\\
\hline
A &   0 & 0 &0   						&   $\bf 1_2$  & 2 &$\bf 2_2$ & 4 & $\bf 2^2_2$ \\
\hline	
\hline		
  D &   0 & +1 & 0   	 						&  $\bf 2$ & $\bf 2$ & $ \bf 2_2$ & 4 & 8 \\
\hline  
  DIII &    -1 & +1 & 1  					&   4  & $\bf 2^2$ & $\bf 2^2$ & $\bf 2^2_2$ & 8 \\
\hline  
  AII &   -1 &  0  	 & 0 					&    $\bf 2_2$ & 4 & $\bf 2^2$  & $\bf 2^2$ & $\bf 2^2_2$\\
\hline  
  CII &   -1 & -1 & 1 	  		&  4 & $\bf 2^2_2$ &  8 & $\bf 2^3$ & $\bf  2^3$  \\
\hline  
  C &  0  & -1 & 0 	 						&   2 & 4  &$\bf 2^2_2$ & 8 & $\bf 2^3$ \\
\hline
  CI &   +1  & -1 & 1  					&   2 & 4 & 8  & $\bf 2^3_2$ & 16\\
\hline 
   AI &   +1 & 0  & 0  					&   $\bf 1_2$ & 2 & 4 & 8 & $\bf 2^3_2$ \\
\hline
   BDI &    +1  & +1 & 1   		&   $\bf 2$ & $\bf 2_2$ & 4 & 8 & 16 \\
\hline
\end{tabular}
\end{center} 
\caption{The table shows the minimum dimension of a Dirac Hamiltonian in the form of \cref{Hform} for each individual symmetry class and spatial dimension. The {\bf bold\/} numbers denote the topological invariant: $\bZ$ or $\bZ_2$. The subscript $2$ indicates that there are two inequivalent representations. Without it, in that symmetry class and dimension there is only one equivalent representation. First, the minimum dimensions of the two complex symmetry classes is from the minimum dimension of the complex representation theory in \cref{complexrep}. Secondly, \cref{size1dc} displays the minimum dimensions of the remaining eight symmetry classes in 1d. By using the interplay between the minimum dimensions of the real symmetry classes and spatial dimensions in \cref{dmove,eight}, the result is shown in this table,  which is in agreement with the indirect result in ref \onlinecite{ribbon_permutation}. 
}
\label{tableminisize}
\end{table}

\subsection{Surface Hamiltonian classification}
	Based on the minimum dimension of the bulk Dirac Hamiltonians, we can regenerate the classification table of topological insulators and superconductors. Similarly, consider the minimum dimension of the gapless surface Hamiltonians in \cref{surfH1}  with a gap opening term 
\begin{equation}
\ch^{\mathrm{surf}}(\vec{k})=m\tilde{\Upsilon}_1+\sum_{i=1}^{d-1}k_i\Upsilon_i,
\end{equation}
where $\tilde{\Upsilon}_1$ and $\Upsilon_i$ have the same anticommutation relation in \cref{gammaanti} as $\tilde{\gamma}_1\rightarrow \tilde{\Upsilon}_1$ and $\gamma_i\rightarrow \Upsilon_i$. The isomorphism between all of the elements of $\ch^{\mathrm{surf}}$ in \cref{elements} and the generators $Cl_{p,q}$ of the real Clifford algebra is simply described by \cref{iso} as $D\rightarrow D-1$ and $d\rightarrow d-1$ due to the absence of $\tilde{\Upsilon}_0$ and $\Upsilon_d$. The relation between the surface and bulk Hamiltonian by \cref{oneplusone} is
\be
\ch^{\mathrm{bulk}}_\#=\ch^{\mathrm{surf}}_\#\otimes Cl_{1,1}.
\ee
In each individual symmetry class and spatial dimension, any surface Hamiltonian in the minimal model is always {\it half\/} the minimum dimension of the bulk Dirac Hamiltonian. Therefore, by considering the minimum dimensions of surface Hamiltonians without and with a gapping term, we can also reproduce the classification table.

\section{Explicit forms of the Dirac Hamiltonians}\label{explicitform}
When spatial dimension $d$ is fixed, there are five nontrivial symmetry classes---``$\bZ,\ \bZ,\ \bZ_2,\ \bZ_2,\ 2\bZ$''. To simplify the following discussion, we define  a complex symmetry class in some dimensions with a $\bZ$ topological invariant as ``$\bZ^C$''. Furthermore, a real symmetry class possessing a $\bZ$ topological invariant is defined as ``$\bZ^R$'', which are called ``primary series'' by Ryu and  Takayanagi\cite{Ryu:2010fk}. Similarly, class ``$2\bZ^R$'' corresponds to $2\bZ$ topological invariant, which indicates the system always has even number of the topological number. For two $\bZ_2$ classes, in the periodic table class ``$\bZ_2^{R,1}$'' is the one right next to class $\bZ^R$ and the remaining $\bZ_2$ class is defined as ``$\bZ_2^{R,2}$''. Class $\bZ_2^{R,1}$ and $\bZ_2^{R,2}$ are also known as the first and second descendant of the primary series. Consider in the same symmetry class and spatial dimension, two different Dirac Hamiltonians with symmetry operators are in an equivalent representation.  According to the presentation theory, there exists an {\it unitary\/} transformation (\cref{equivalent} will show the meaning of the transformation) from one Hamiltonian and the corresponding symmetry operators to the others. Therefore, to construct one Hamiltonian for each individual symmetry and spatial dimension, we only need to find out one Hamiltonian for each inequivalent representation.

In \cref{tableminisize}, we observe that in even spatial dimension, say $d=2n$, the {\it minimum dimension\/} relation of a Dirac Hamiltonian in the five non-trivial symmetry classes is 
\begin{align}
&2size(\bZ^C)=2size(\bZ^R)=size(\bZ_2^{R,1}) \nonumber \\
=&size(\bZ_2^{R,2})=size(2\bZ^R)=2^{n+1}  \label{evensize}
\end{align}
In odd spatial dimension, say $d=2n-1$, the relation is slightly different from in even dimension
\begin{align}
&2size(\bZ^C)=2size(\bZ^R)=2size(\bZ_2^{R,1}) \nonumber \\
=&size(\bZ_2^{R,2})=size(2\bZ^R)=2^{n+1} \label{oddsize}
\end{align}
In the real symmetry classes, we use the explicit matrix of a Dirac Hamiltonian for class $\bZ^R$ to construct the matrix expression for class $\bZ_2^{R,1}$, $\bZ_2^{R,2}$, and $2\bZ^R$. When we observe the periodic \cref{table:topo}, the relation of time reversal and particle hole operators can be found in odd(even) dimension. 
\be
(T^{d}_{\bZ^R})^2=-(T^{d}_{2\bZ^R})^2=(C^{d}_{\bZ_2^{R,1}})^2=(C^{d+2}_{\bZ^R})^2=(T^{d=\mathrm{even}}_{\bZ_2^{R,1}})^2 \label{Trelation}
\ee
and(or)
\be
(C^{d}_{\bZ^R})^2=-(C^{d}_{2\bZ^R})^2=-(T^{d}_{\bZ_2^{R,1}})^2=-(T^{d+2}_{\bZ^R})^2=(C^{d=\mathrm{even}}_{\bZ_2^{R,1}})^2 \label{Crelation}
\ee
The last terms in both of the equations are only for even spatial dimension. Here the superscript indicates spatial dimension and the subscript indicates a topological invariant class. If the explicit expressions of class $\bZ^R$ is known, Dirac Hamiltonians with two inequivalent representations in odd(even) dimension are written as
\be
\ch({\bf k})_{\bZ^R}=\pm M\tilde{\gamma}_0+\sum_{i=1}^dk_i\gamma_i. \label{Hamform}
\ee
The minus and plus signs represent two inequivalent representations. This Hamiltonian preserves TRS and(or) particle PHS with time reversal operator $(T_{\bZ^R})$ and(or) particle-hole operator $(C_{\bZ^R})$. 

\subsection{Two parent Hamiltonians}
To construct Dirac Hamiltonians for all topological invariant classes, we start from writing down Dirac Hamiltonians for class BDI in 1-d, corresponding to class $\bZ^R$:
\be
\ch=\pm m\sigma_x + k_x\sigma_y \label{parent1d}
\ee
with time reversal operator $T=K$ and particle hole operator $C=\sigma_z K$. The signs indicate two inequivalent representations. Also, we build a Dirac Hamiltonian for class D in 2-d, corresponding to class $\bZ^R$:
\be
\ch=\pm m\sigma_y + k_x\sigma_x+k_y\sigma_z
\ee
with particle hole operator $C=K$. The former is the base Hamiltonian to construct all other Dirac Hamiltonians in odd dimension. Likewise, the latter is for even dimension. 

\section{Different classes, the same dimension }
To construct other Dirac Hamiltonians, we have to separate to two cases. First, we will fix the dimension and build a Dirac Hamiltonian in different topological invariant classes. Secondly, we will construct a Dirac Hamiltonian in class $\bZ^R$ in $d+2$ dimension from the same class in $d$ dimension. Thus, a Dirac Hamiltonian for any class and in any dimension can be built from these two rules. Some parts of the explicit forms of Dirac Hamiltonians are the similar with Teo and Kane's results\cite{Teo:2010fk}.

\subsubsection{$2\bZ^R$ in odd(even) dimension}
The construction of a Dirac Hamiltonian in class $2\bZ^R$ is from class $\bZ^R$ in the same spatial dimension.  By \cref{evensize,oddsize}, the minimum dimension of a Dirac Hamiltonian in class $2\bZ^R$ is twice as big as in class $\bZ^R$. Moreover, as the spatial dimension is fixed, $T_{2\bZ^R}^2$ and(or) $C_{2\bZ^R}^2$ have different signs from $T_{\bZ^R}^2$ and(or) $C_{\bZ^R}^2$. To have these two symmetries, we write down the explicit matrix form for class $2\bZ^R$ with two inequivalent representations from class $\bZ^R$ as 
\be
\ch({\bf k})_{2\bZ^R}=(\pm M\tilde{\gamma}_0+\sum_{i=1}^dk_i\gamma_i)\otimes \bI
\ee
When the symmetry operators are defined as $T_{2\bZ^R}=T_{\bZ^R}\otimes \sigma_y$ and(or) $C_{2\bZ^R}=C_{\bZ^R}\otimes \sigma_y$, the corresponding symmetries of $2\bZ^R$ are preserved.

\subsubsection{$\bZ_2^{R,1}$ in odd dimension}
Consider a Dirac Hamiltonian for class $\bZ_2^{R,1}$ in odd dimension. The minimum dimension of a Dirac Hamiltonian is the same with class $\bZ^R$. Therefore, the explicit form is 
\be
\ch({\bf k})_{\bZ_2^{R,1}}=\ch({\bf k})_{\bZ^R}.
\ee  
To have a Hamiltonian belonging to class $\bZ_2^{R,1}$, we only need to keep one of those two symmetries from class $\bZ^R$. That is, as $d=4l+1$ and $d=4l+3$, TRS and PHS are kept respectively. It should be noticed that although class $\bZ^R$ and class $\bZ_2^{R,1}$ have the same form of the Hamiltonian, the restrictions of perturbations added into the Hamiltonian for class $\bZ^R$ is more rigorous than for class $\bZ_2^{R,1}$.  The reason is that in odd dimension class $\bZ^R$ preserves two symmetries but class $\bZ_2^{R,1}$ preserves one. In the case of a single copy of the Hamiltonian, the gap opening term for both of the classes is not allowed by the symmetries. However, the difference of the restrictions is shown from the two copies of the Hamiltonian. For class $\bZ^R$ the gap opening term is still forbidden; for class $\bZ_2^{R,1}$ it becomes allowed. That is the reason that the two classes possess different topological invariants: $\bZ$ and $\bZ_2$.

\subsubsection{$\bZ_2^{R,2}$ in odd dimension}
The relation of the symmetry operator between class $\bZ^R$ and $\bZ_2^{R,1}$ in \cref{Trelation,Crelation} shows that $T_{\bZ^R}^2$ and $C_{\bZ_{2R}}^2$ have the same sign but $C_{\bZ^R}^2$ and $T_{\bZ_{2R}}^2$ have the different sign. Furthermore, a Dirac Hamiltonian in class $\bZ_2^{R,2}$ is twice the minimum dimension in class $\bZ^R$. Therefore, we define the symmetry operators in class $\bZ_2^{R,2}$ as  $T_{\bZ_2^{R,2}}=C_{\bZ^R}\otimes \sigma_y$ and $C_{\bZ_2^{R,2}}=T_{\bZ}\otimes \sigma_x$. At the same time, the Hamiltonian must be defined as 
\be
\ch({\bf k})_{\bZ_2^{R,2}}=\ch({\bf k})_{\bZ^R}\otimes\sigma_z. 
\ee

\subsubsection{$\bZ_2^{R,1}$ in even dimension}
A system in class $\bZ^R$ in even dimension has only one of the two symmetries ($T_{\bZ^R}$ and $C_{\bZ^R}$). However, a system in class $\bZ_2^{R,1}$ possesses those two symmetries. Therefore, to construct the symmetry preserving Hamiltonian in class $\bZ_2^{R,1}$, we have to enlarge the original Hamiltonian of class $\bZ^R$. This is the reason that for $\bZ_2^{R,1}$ the minimum dimension of the Hamiltonian in even dimension is twice as big as for $\bZ^R$. The expression of the Hamiltonian is written as
\be
\ch({\bf k})_{\bZ_2^{R,1}}=(\pm M\tilde{\gamma}_0+\sum_{i=1}^dk_i\gamma_i)\otimes \sigma_z
\ee
Here $\sigma_z$ keeps the Hamiltonian preserving TRS and PHS. To verify this, we need to find those two corresponding symmetry operators. There are two situations from the symmetry of class $\bZ^R$ to define time reversal operator and particle hole operator for class $\bZ_2^{R,1}$. For the first case, as $d=4l$, a system of $\bZ^R$ only has TRS $(T_{\bZ^R})$. To preserving TRS and PHS for class $\bZ_2^{R,1}$, the two symmetry operators can be defined as $T_{\bZ_2^{R,1}}=T_{\bZ^R}\otimes \bI$ and $C_{\bZ_2^{R,1}}=T_{\bZ^R}\otimes \sigma_x$; the definition obeys the symmetry operator relation in \cref{Trelation}. For the other case, as $d=4l+2$ there is only particle hole symmetry $C_{\bZ^R}$ in class $\bZ^R$. Similarly, the two symmetry are defined as $T_{\bZ_2^{R,1}}=C_{\bZ^R}\otimes \sigma_y$ and $C_{\bZ_2^{R,1}}=C_{\bZ^R}\otimes \bI$, which obey the relation in \cref{Crelation}.

\subsubsection{$\bZ_2^{R,2}$ in even dimension}
In even dimension a Dirac Hamiltonian in class $\bZ_2^{R,2}$ preserves one of the two symmetries in class $\bZ_2^{R,1}$. By \cref{evensize}, the minimum dimensions of $H({\bf k})_{\bZ_2^{R,2}}$ and $H({\bf k})_{\bZ_2^{R,1}}$ are the same. This situation is similar with the construction of a Hamiltonian in class $\bZ_2^{R,1}$ in odd dimension. Therefore, in this case 
\be
\ch({\bf k})_{\bZ_2^{R,2}}=\ch({\bf k})_{\bZ_2^{R,1}}. 
\ee
The corresponding symmetry in different dimensions can be considered separately: $d=4l$ and $d=4l+2$. For the former, $C_{\bZ_2^{R,1}}=T_{\bZ^R}\otimes \sigma_x$; for the latter, $T_{\bZ_2^{R,1}}=C_{\bZ^R}\otimes \sigma_y$.

\begin{widetext}

\begin{table} 
\begin{tabular}{|c||c|c|c|c|} 
\hline
Dimension & Odd & Symmetry Operators & Even  & Symmetry Operators \\
\hline
\hline
$\bZ^C$ & $\ch_{\bZ^R}$ & $S_{\bZ^C}=T_{\bZ^R}C_{\bZ^R}$ &$\ch_{\bZ^R}$ & $0$  \\
\hline
$\bZ^R$ & $\ch_{\bZ^R}$ & $T_{\bZ^R}$ and $C_{\bZ^R}$ &$\ch_{\bZ^R}$   & $T_{\bZ^R}$ as $d=4l$ or $C_{\bZ^R}$ as $d=4l+2$\\	
\hline
\multirow{2}{*}{$\bZ_2^{R,1}$} & \multirow{2}{*}{$\ch_{\bZ^R}$} & $C_{\bZ_2^{R,1}}=C_{\bZ^R}$ as $d=4l+1$  & \multirow{2}{*}{$\ch_{\bZ^R}\otimes\sigma_z$} & $T_{\bZ_2^{R,1}}=T_{\bZ^R}\otimes \bI$ and $C_{\bZ_2^{R,1}}=T_{\bZ^R}\otimes \sigma_x$  as $d=4l$ or \\ 
 & & $T_{\bZ_2^{R,1}}=T_{\bZ^R}$ as $d=4l+3$ & & $T_{\bZ_2^{R,1}}=C_{\bZ^R}\otimes \sigma_y$ and $C_{\bZ_2^{R,1}}=C_{\bZ^R}\otimes \bI$ as $d=4l+2$ \\
\hline
\multirow{2}{*}{$\bZ_2^{R,2}$} & \multirow{2}{*}{$\ch_{\bZ^R}\otimes \sigma_z$} & \multirow{2}{*}{$T_{\bZ_2^{R,2}}=C_{\bZ^R}\otimes \sigma_y$ and $C_{\bZ_2^{R,2}}=T_{\bZ^R}\otimes \sigma_x$}  & \multirow{2}{*}{$\ch_{\bZ^R}\otimes\sigma_z$} & $C_{\bZ_2^{R,2}}=T_{\bZ^R}\otimes \sigma_x$ as $d=4l$ or  \\ & & & & $T_{\bZ_2^{R,2}}=C_{\bZ^R}\otimes \sigma_y$ as $d=4l+2$ \\
\hline
\multirow{2}{*}{$2\bZ^R$} & \multirow{2}{*}{$\ch_{\bZ^R}\otimes \bI$} & \multirow{2}{*}{$T_{2\bZ^R}=T_{\bZ^R}\otimes\sigma_y$, $C_{2\bZ^R}=C_{\bZ^R}\otimes\sigma_y$} & \multirow{2}{*}{$\ch_{\bZ^R}\otimes \bI$}  & $T_{2\bZ^R}=T_{\bZ^R}\otimes\sigma_y$ as $d=4l$  \\
& &  & & $C_{2\bZ^R}=C_{\bZ^R}\otimes\sigma_y$  as $d=4l+2$ \\
\hline
\end{tabular} 
\caption{In any spatial dimension, the construction of minimal-size Dirac Hamiltonians for five non-trivial symmetry classes from $H_{\bZ^R}$ is shown in the table.} \label{expressionHbulk}
\end{table}

\end{widetext}

\subsection{The same class, different dimensions} \label{sameclass}
	Suppose we know the explicit form of a Dirac Hamiltonian $\ch({\bf k})_{\bZ^R}^{d}$ in $d$ dimension. By \cref{evensize,oddsize}, a Dirac Hamiltonian $\ch({\bf k})_{\bZ^R}^{d+2}$ in $d+2$ dimension is twice the minimum dimension of $\ch({\bf k})_{\bZ^R}^{d}$. To construct this higher dimension Hamiltonian, we need to have two extra linear momentum terms anticommuting with the enlarged $H({\bf k})_{\bZ^R}^{d}$. To satisfy this relation, the new Hamiltonian is constructed from the tensor product of the three Pauli matrices
\be
\ch({\bf k})_{\bZ^R}^{d+2}=\ch({\bf k})_{\bZ^R}^{d}\otimes \sigma_y+k_{d+1}\bI\otimes\sigma_x+k_{d+2}\bI\otimes \sigma_z. \label{2+dZR}
\ee	
Furthermore, from the dimension independent symmetry operator relations of \cref{Trelation,Crelation}, $(T_{\bZ^R}^{d+2})^2$ and $(C_{\bZ^R}^{d})^2$ have different signs and $(C_{\bZ^R}^{d+2})^2$ and $(T_{\bZ^R}^{d})^2$ have the same sign. To obey these relations and the symmetry equations, the symmetry operators should be 
\be
T_{\bZ^R}^{d+2}=C_{\bZ^R}^{d}\otimes\sigma_y ,\ C_{\bZ^R}^{d+2}=T_{\bZ^R}^{d}\otimes \bI
\ee
As $d$ is odd, the Hamiltonian preserves both of the symmetries. As $d=4l(d=4l+2)$, the former(latter) operator is kept. }

	There is an interesting discussion about symmetry breaking as we add $T_{\bZ^R}C_{\bZ^R}$ into the Hamiltonian, which will be used in the last two sections. In class $\bZ^R$ in odd dimension the Hamiltonian preserve TRS and PHS. Recalling the parent Hamiltonian in 1-d in \cref{parent1d}, $T_{\bZ^R}^1C_{\bZ^R}^1=\sigma_z$ is hermitian and breaks PHS but preserves TRS. We apply the relation in \cref{2+dZR}, to find the generalization of $T_{\bZ^R}C_{\bZ^R}$ breaking symmetry in any odd dimension. Therefore, such generalization is that as $d=4l+1(4l+3)$, $T_{\bZ^R}C_{\bZ^R}$ breaks PHS(TRS) and preserves TRS(PHS).

{ 

\section{The Dirac Hamiltonian for $\bZ_2$ 3D topological insulators}\label{3dTI}

We build a concrete lattice model of a Dirac Hamiltonian, to demonstrate gapless surface modes for 3D topological insulators--- a strong topological insulator has a single Dirac cone on the surface and a weak topological insulator has two Dirac cones on the surface. One\cite{Qi:2008sf} of the simplest lattice models is written as
\begin{align}
\ch=&\sum_{m,n,l}[M\tilde{\gamma}_0c^\dagger_{m,n,l}c_{m,n,l} \nonumber \\
&+\frac{\tilde{\gamma}_0+i\gamma_1}{2}c^\dagger_{m,n,l}c_{m+1,n,l}+\frac{\tilde{\gamma}_0-i\gamma_1}{2}c^\dagger_{m,n,l}c_{m-1,n,l}\nonumber \\
&+\frac{\tilde{\gamma}_0+i\gamma_2}{2}c^\dagger_{m,n,l}c_{m,n+1,l}+\frac{\tilde{\gamma}_0-i\gamma_2}{2}c^\dagger_{m,n,l}c_{m,n-1,l}\nonumber \\
&+\frac{\tilde{\gamma}_0+i\gamma_3}{2}c^\dagger_{m,n,l}c_{m,n,l+1}+\frac{\tilde{\gamma}_0-i\gamma_3}{2}c^\dagger_{m,n,l}c_{m,n,l-1}]. \label{3DH}
\end{align}
For simplicity, we let the lattice constant be 1. These gamma matrices have the explicit form of the tensor products of two Pauli matrices corresponding to orbital and spin degrees of freedom respectively: $\gamma_1=\sigma_z\otimes\sigma_x,$  
$\gamma_2=\sigma_z\otimes\sigma_y,$ 
$\gamma_3=\sigma_z\otimes\sigma_z,$ 
$\tilde{\gamma}_0=\sigma_x \otimes \bI$. However, without knowing this expression of the gamma matrices we still can use anticommutation relation of the gamma metrics to derive the surface states. We only need to know that the minimum size of the Dirac Hamiltonian is $4\times 4$ from \cref{tableminisize}. It is easy to check that the Hamiltonian preserves TRS with the time reversal operator $T=\bI\otimes \sigma_yK$. To determine the topological invariants, we consider the Hamiltonian in momentum space:
\begin{align}
\ch_{\rm{TI}}=&\sum_{\bf p}c^\dagger_{\bf p}\big [(M+\cos p_x+\cos p_y+\cos p_z)\tilde{\gamma}_0+\sin p_x\gamma_1 \nonumber \\
&+ \sin p_y \gamma_2 +\sin p_z \gamma_3\big ]c_{\bf p}. \label{3dstr}
\end{align}
From this Hamiltonian, we can see that quantum phase transition points are at $M=3,\ 1,\ -1,\ -3$. The system is also preserved by inversion symmetry with the operator $P=\tilde{\gamma}_0$. Fu and Kane's paper\cite{Fu:2007uq} shows that the corresponding topological invariants are
\begin{align*}
&M>3 		& (0;&000)  \\
3>&M>1		& (1;&000)  \\
1>&M>-1		& (0;&111)  \\
-1>&M>-3		& (1;&111)  \\
-3>&M		& (0;&000)
\end{align*}
Consider open boundary condition in the z direction and periodic boundary conditions in the other directions. Momentum $p_x$ and $p_y$ are still good quantum numbers in the Dirac Hamiltonian  
\begin{align}
\ch_{\rm{TI}}=&\sum_{p_x,p_y,l}\Big\{ c^\dagger_{p_x,p_y,l} \big [(M+\cos p_x+\cos p_y)\tilde{\gamma}_0+\sin p_x\gamma_1  \nonumber \\
&+\sin p_y\gamma_2\big ] c_{p_x,p_y,l} +c^\dagger_{p_x,p_y,l} \frac{\tilde{\gamma}_0+i\gamma_3}{2}c_{p_x,p_y,l+1}  \nonumber \\
&+c^\dagger_{p_x,p_y,l} \frac{\tilde{\gamma}_0-i\gamma_3}{2}c_{p_x,p_y,l-1} \Big \}
\end{align}
To solve the Schr\"{o}dinger equation, let $\sum_l\alpha_lc^\dagger_l\mid 0\rangle$ be an eigenstate. We can write down a recurrence equation, which is the Harper equation 
\begin{align}
E_\mu\alpha_l&=[(M+\cos p_x +\cos p_y)\tilde{\gamma}_0 + \sin p_x \gamma_1 +\sin p_y \gamma_2]\alpha_l \nonumber \\
&+\frac{\tilde{\gamma}_0+i\gamma_3}{2}\alpha_{l+1}+\frac{\tilde{\gamma}_0-i\gamma_3}{2}\alpha_{l-1} \label{Harper3d}
\end{align}
To simplify the calculation, define $\ket{+}$ and $\ket{-}$, two subspaces, which satisfy
\begin{align}
\frac{\mathbb{I}+i\tilde{\gamma}_0\gamma_3}{2}\ket{+} &=\ket{+}, & \frac{\mathbb{I}-i\tilde{\gamma}_0\gamma_3}{2}\ket{+} &=0  \\
\frac{\mathbb{I}-i\tilde{\gamma}_0\gamma_3}{2}\ket{-} &=\ket{-} , & \frac{\mathbb{I}+i\tilde{\gamma}_0\gamma_3}{2}\ket{-} &=0,
\end{align}
That is, $\ket{+}$ and $\ket{-}$ are eigenspaces of $i\tilde{\gamma}_0\gamma_3$ with eigenvalues $\pm 1$. Therefore, $\alpha_l$ can be decomposed to $\alpha_{l}^+\ket{+} +\alpha_l^-\ket{-}$. Rewrite \cref{Harper3d}
\begin{align}
& E_\mu(\alpha_{l}^+\mid + \rangle+\alpha_l^-\ket{-}) \nonumber \\
=&(M+\cos p_x +\cos p_y)(\alpha_{l}^+\tilde{\gamma}_0\ket{+}+\alpha_l^-\tilde{\gamma}_0\ket{-}) \nonumber \\
&+(\sin p_x \gamma_1 +\sin p_y \gamma_2)(\alpha_{l}^+\ket{+} +\alpha_l^-\ket{-}) \nonumber \\
&+\alpha_{l+1}^+\tilde{\gamma}_0\ket{+}
+\alpha_{l-1}^-\tilde{\gamma}_0\ket{-}
\end{align}
To capture the surface states with zero energy ($E_\mu=0$), let $\sin p_x,\ \sin p_y$ vanish and will be recovered later. Therefore, we have two recurrence equations by comparing with the coefficients of $\tilde{\gamma}_0\ket{+}$ and $\tilde{\gamma}_0\ket{-}$
\begin{align}
-(M+\cos p_x + \cos p_y)\alpha_l^\pm=\alpha_{l\pm1}^\pm
\end{align}
For the open boundary condition in the z direction, the wave function must vanish at the boundary: $\alpha_0^\pm=0$ and $\alpha_{L+1}^\pm=0$. To obtain surface states, we have $\alpha^+_l\rightarrow 0$ as $l\rightarrow L$ and $\alpha^-_l\rightarrow 0$ as $l\rightarrow 0$ when $|M+\cos p_x +\cos p_y |<1$. Since $\alpha^\pm_l$ has two degrees of freedom, if the surface states exist, there are at least two orthonormal surface states ($\ket{+_{1,2}}$ or $\ket{-_{1,2}}$) on each side. The number of the surface states for different $M$ is given by
\begin{align*}
|M|&>3 &  &\text{No surface states}  \\
3>M&>1 & & 1\times2\text{ surface states as }{\bf p}=(\pi,\pi) \\
1>M&>-1 & & 2\times2 \text{ surface states as }{\bf p}=(\pi,0) \text{, } (0,\pi) \\
-1>M&>-3 & & 1\times2\text{ surface state as }{\bf p}=(0,0).
\end{align*}
After considering the zero energy case above, we recover $\sin p_x$ and $\sin p_y$ in the Hamiltonian in \cref{Harper3d}. Since $i\tilde{\gamma}_0\gamma_3$ commutes with $\gamma_1$ and $\gamma_2$, the surface Hamiltonian can be written as the projection of \cref{Harper3d} on $\ket{+ _{1,2}}$ or $\ket{-_{1,2}}$ on each surface. To simplify the problem, we focus on one surface, say $\ket{+ _{1,2}}$. The expression of the surface Hamiltonians for different $M$ is 
\begin{small}
\begin{align}
3>M&>1 & \ch^{\mathrm{surf}} &=\sin \Delta p_x \Upsilon_1+\sin \Delta p_y \Upsilon_2 \\
1>M&>-1 & \ch^{\mathrm{surf}}&=\sin \Delta p_x \sigma_z\otimes \Upsilon_1 - \sin \Delta p_y \sigma_z \otimes \Upsilon_2  \label{weakregion}  \\
-1>M&>-3 & \ch^{\mathrm{surf}}&=-\sin \Delta p_x \Upsilon_1 - \sin \Delta p_y \Upsilon_2,
\end{align}
\end{small}
where $\Upsilon_{i\alpha\beta}=\bra{+_\alpha,\vec{p}} \gamma_i \ket{+_\beta,\vec{p}}$, $\{\Upsilon_i,\Upsilon_j\}=2\delta_{ij}\bI$, and $\Delta p=\vec{p}-{\bf p}$. The Pauli matrix $\sigma_z$ of \cref{weakregion} is in the basis near $\vec{p}=(0,\pi)$ and $(\pi,0)$. The momentum ${\bf p}$ is for the entire first Brillouin zone but in the region so that $|M+\cos p_x +\cos p_y |<1$. We note that these surface Hamiltonian still preserve TRS due to time reversal preserved $i\tilde{\gamma}_0\gamma_3$. $\Upsilon_1$ and $\Upsilon_2$ anticommute with each other because of the equivalent relation 
\be
(\frac{\bI+i\gamma_1\gamma_3}{2})\gamma_i(\frac{\bI+i\gamma_1\gamma_3}{2}) \sim \begin{pmatrix} \Upsilon_i & 0 \\ 0 & 0 \end{pmatrix}\ {\text for}\ i=1,2 
\ee
and anticommutation relation between $\gamma_1$ and $\gamma_2$. For a strong topological insulator, if we wish to open a surface gap, TRS must be broken. For a weak topological insulator, we will show later, some density wave can open up the surface gap.

\subsection{The gapped surface state of a strong topological insulator} \label{strong surface}
For a strong topological insulator, we are interested in the physics of the surface state gapped by magnetic materials. These magnetic materials, having the opposite magnetization directions, form a domain wall. We will show on the surface around this domain wall there is a chiral gapless state. We can keep the linear part of $\bf p$ in the surface Hamiltonian for a strong topological insulator, say $3>M>1$, to capture the lowest energy physics
\be 
\ch^{\mathrm{surf}}=p_x\Upsilon_1+p_y\Upsilon_2+m\Upsilon_3,
\ee 
where $m\Upsilon_3$ from the magnetic materials breaks TRS. To have a domain wall we can set up $m>0$ as $x>0$, and $m<0$ as $x<0$; momentum $p_x$ is not a good quantum number so let $p_x\rightarrow -i\partial/\partial x$. After following the similar calculation in \cref{surfacecal}, $H^{\mathrm{surf}}$ is a $2\times 2$ matrix from the projection of the $4\times 4$ bulk Hamiltonian matrix.
We find only one eigenstate with an energy
\be
\Psi_{FM}=e^{-\int_0^xm(x')dx'}\psi,\ E=\pm p_y,
\ee
where $\psi$ is an eigenstate of $i\Upsilon_1\Upsilon_3$ with an eigenvalue $1$. Since $i\Upsilon_1\Upsilon_3$ commute with $\Upsilon_2$, $\psi$ is chosen as an eigenstate of $\Upsilon_2$ with an eigenvalue $1$ or $-1$, which determines the sign of $E$. There is a chiral gapless edge state around $x=0$. We can argue that the regions of $x>0$ and $x<0$ demonstrate a half quantum Hall effect. Quantum Hall effect can be described by the TKNN number\cite{Thouless:1982rz}. Those regions have the same absolute values of TKNN number because the sign of $m$ does not change physics. Moreover, a chiral edge state between those regions shows the difference of the TKNN numbers is one. Thus, the surface of a strong topological insulator gapped by a uniform magnetic material exhibits a half quantum Hall effect. 

\subsection{The gapped surface state of a weak topological insulator} \label{gapWTI}
A weak topological insulator with zero strong index have two Dirac cones on the surface. When translational symmetry is broken by adding proper CDW, the scattering of these two Dirac cones opens gaps. Later, we will show that helical gapless edge states\cite{Liu:2011fk} are seen around this CDW domain wall, which is similar with the domain wall of magnetic materials. Consider a CDW creates potential difference, preserving TRS, for each nearest neighbor side along the x and y directions
\begin{align}
\ch_{\mathrm{CDW}}&=\epsilon \sum (-1)^{n+m}c^\dagger_{n,m,l} c_{n,m,l}
\end{align}
Transform this additional Hamiltonian to the momentum space 
\be
\ch_{\mathrm{CDW}}=\epsilon \sum c^\dagger_{p_x+\pi,p_y+\pi,l}c_{p_x,p_y,l}
\end{equation}
We focus on the weak topological insulator phase in the region $1>M>-1$. The surface Hamiltonian in \cref{weakregion} with CDW in the basis of the four surface states $\ket{+_{1,2},0,\pi},\ \ket{+_{1,2},\pi,0}$ is expressed as
\be 
\ch_{\mathrm{CDW}}^{\mathrm{surf}}=\sin \Delta p_x \sigma_z\otimes \Upsilon_1 - \sin\Delta p_y \sigma_z \otimes \Upsilon_2  +\epsilon \sigma_x\otimes \bI
\ee
This CDW is the only term opening a gap on the surface states and preserving TRS. However, a question arises: what is the physical property of this surface system? Consider the low energy physics--- $\sin \Delta p_x\rightarrow \Delta p_x$ and $\sin \Delta p_y \rightarrow \Delta p_y$. Again we set up a domain wall along the x direction: $\epsilon>0$ as $x>0$, and $\epsilon<0$ as $x<0$. The momentum $p_x$ is not a proper quantum number so $p_x\rightarrow -i\partial/\partial x$. After solving the Schr\"{o}dinger equation, the two eigenstates and the corresponding eigenvalues are given by 
\begin{align}
\Phi_{CDW\pm}=e^{-\int_0^x\epsilon(x')dx'}
\varphi_\pm
,\ E_{\pm}=\pm p_y,
\end{align}
where $\varphi_\pm$ is an eigenstate of $\sigma_x\otimes\Upsilon_1$ and $\sigma_z\otimes\Upsilon_2$ with the eigenvalues $-1$ and $\pm 1$ respectively because $\sigma_x\otimes\Upsilon_1$ commutes with $\sigma_z\otimes\Upsilon_2$. These two states $\Phi_{CDW\pm}$ provide helical gapless spectrum around the domain wall. We can make a similar argument with a half quantum Hall effect that the states on the surface gapped by a uniform CDW have a half quantum spin Hall effect.

\section{Generalization of a non-trivial surface on a {\it strong\/} topological insulator}\label{strongreduction}
The preceding section shows that on a 3D topological insulator the gapless surface states gapped by TRS breaking exhibits a half quantum Hall effect. However, this example is a part of  the big scheme: when the protected gapless surface states is gapped by some symmetry breaking, this surface system, belonging to a different $d-1$ spatial dimension symmetry class, might possess nontrivial topological properties. First, we consider three symmetry breaking cases: time reversal, particle-hole, and chiral symmetries. Although PHS breaking is not physical, in our mathematical generalization PHS plays the essential role as TRS. To have nontrivial topological physics, the surface states in (d-1)-dimension with symmetry breaking must fall into class $\bZ$ or $\bZ_2$ because the gapless edge states of the surface are not protected in class $``0"$. Thirdly, for $\bZ$ and $\bZ_2$ topological insulators and superconductors, Dirac cones, representing the gapless states, on the surfaces are not necessarily sitting at symmetry points. In this section, to avoid complexity of scattering between Dirac cones with different momenta, we focus on those Dirac cones at $\vec{0}$ in Brillouin zone. That is, the surface Hamiltonian for class $\bZ^R$ is expressed as
\be 
\ch^{\mathrm{surf}}_{\bZ^R}=\sum_{i=1}^{d-1}k_i\Upsilon_i,
\ee
where $\{\Upsilon_i,\Upsilon_j\}=2\delta_{ij}\dI$. This Hamiltonian $\ch^{\mathrm{surf}}_{\bZ^R}$ is the projection of the bulk Hamiltonian $H_{\bZ^R}$ in \cref{Hamform}. By knowing \cref{expressionHbulk}, any surface Hamiltonian in nontrivial symmetry class can be generated by $\ch^{\mathrm{surf}}_{\bZ^R}$. Fourthly, we expect that once the gapless surface states are gapped, the surface always exhibits a nontrivial topological behavior. However, there are two exceptions: $\bZ^C\leftarrow \bZ_2^{R,2}$ and $\bZ^C\leftarrow 2\bZ^R$ from the bulk in odd dimension to the surface in even dimension. That is, the surface states may be gapped by different forms of the same kind symmetry breaking terms. The surfaces of these two systems might be in a nontrivial or trivial phase. In the following, we simplify our focus on the cases that the gapless surface states only can be gapped by a unique gapping term in that new symmetry class. Such gapped surface states inevitably stay in a nontrivial phase.

By observing the classification \cref{table:topo}, we find that there are eight ways to have nontrivial topological physics on the surface by breaking one symmetry and only one way by breaking two symmetries. Such paths of breaking symmetries can be classified to four categories as a nontrivial topological class in $d$ bulk dimension changes to another in $d-1$ surface dimension: $\bZ\leftarrow \bZ,\ \bZ_2\leftarrow \bZ,$ $\bZ\leftarrow \bZ_2,$ and $\bZ_2\leftarrow \bZ_2$. In the following subsection, we discuss these four categories in details.

\subsubsection{$\bZ\leftarrow \bZ$}\label{bZbZ}
In odd dimension $d=2n+1$, there are three nontrivial physics cases of the surface Hamiltonian by breaking one symmetry: $\bZ^C\leftarrow \bZ^C$, $\bZ^R\leftarrow\bZ^R$, and $2\bZ^R\leftarrow2\bZ^R$. For the first case, for the minimum model $S_{\bZ^C}$ chiral symmetry operator is hermitian by choosing a proper phase. We can project $S_{\bZ^C}$ into the surface as half-size $S_{\bZ^C}^\mathrm{surf}$ breaking chiral symmetry. The surface states fall into class $\bZ^C$ in 2n-dimension. For the two remaining cases ($\bZ^R$ and $2\bZ^R$), chiral symmetry operator is the combination of TRS and PHS operators $S=TC$. From the result of \cref{sameclass}, $S_{\bZ^R}$ and $S_{2\bZ^R}=S_{\bZ^R}\otimes\bI$ break TRS(PHS) and keeps PHS(TRS) as $d=4l+3(4l+1)$. Hence, after adding the surface projection of $T_{\bZ^R}C_{\bZ^R}$ to the surface states, topological invariant class $\bZ^R$ and $2\bZ^R$ are still unchanged in the symmetry breaking surface. Suppose there are a Dirac cone for $\bZ^C$ and $\bZ^R$ and a pair of Dirac cones for $2\bZ^R$ on the surfaces; the gapped surface Hamiltonians for those three cases are written as 
\begin{align}
\ch^{\mathrm{surf}}_{\bZ^C\leftarrow\bZ^C}=&mS_{\bZ^C}^\mathrm{surf}+\sum_{i=1}^{2n}k_i\Upsilon_i,  \\
\ch^{\mathrm{surf}}_{\bZ^R\leftarrow\bZ^R}=&mS_{\bZ^R}^\mathrm{surf}+\sum_{i=1}^{2n}k_i\Upsilon_i,  \\
\ch^{\mathrm{surf}}_{2\bZ^R\leftarrow2\bZ^R}=&mS_{\bZ^R}^\mathrm{surf}\otimes\bI+\sum_{i=1}^{2n}k_i\Upsilon_i \otimes\bI.
\end{align}
We note that the minimum matrix dimension of the $2n$-dimensional surface Hamiltonians $H_{\bZ^C}$, $H_{\bZ^R}$, $H_{2\bZ^R}$ are $2^n\times 2^n$, $2^n\times 2^n$, and $2^{n+1}\times 2^{n+1}$ from the matrix dimension reduction of  the bulk Dirac Hamiltonians in $2n+1$ dimensions. On the other hand, the bulk Dirac Hamiltonians in $2n$ dimensions for class $\bZ^C$, $\bZ^R$, and $2\bZ^R$ have the same minimum matrix dimensions of those surface Hamiltonians respectively. Therefore, in class $\bZ$ the chiral symmetry operator is the only symmetry preserving mass term in the surface Hamiltonians. By following the similar argument in \cref{strong surface}, the surface Hamiltonian exhibits half topological physics. That is, the topological numbers of $H_{\bZ^C}^{\mathrm{surf}}$ and $H_{\bZ^R}^{\mathrm{surf}}$ is $\pm 1/2$ and the topological number of $H_{2\bZ^R}^{\mathrm{surf}}$ is $\pm 1$, which is different from even topological number in class $2\bZ$. 


Consider two symmetries are broken. We find that only in odd dimension $\bZ^C\leftarrow \bZ^R$ the surface Hamiltonian has a non-trivial topological physics. The surface Hamiltonian $\ch^{\mathrm{surf}}_{\bZ^C\leftarrow \bZ^R}$ is identical to $\ch^{\mathrm{surf}}_{\bZ^R\leftarrow \bZ^R}$. The mass term $S^{\mathrm{surf}}_{\bZ^C}$ breaks one of the symmetries. The other symmetry breaking does not anticommute with all of the elements of the Hamiltonian. The reason is that another anticommuting term causes that the system is always in the trivial phase; it contradicts class $\bZ^C$ in even dimension has the only one symmetry preserving mass term in the minimal Hamiltonian. Therefore, this symmetry breaking does not change the topological invariant $\bZ$ but switches the symmetry class from $\bZ^R$ to $\bZ^C$. 

When we investigate the surface Hamiltonian in even dimension $d=2n$ after one symmetry breaking, there is no nontrivial topological physics in this case.  

\subsubsection{$\bZ_2\leftarrow\bZ$}
The only way from $\bZ$ to $\bZ_2$ by breaking a symmetry is from in odd dimension $(d=2n+1)$ $\bZ^R$ to $\bZ_2^{R,2}$. The periodic table shows that to change this topological property we require TRS breaking as $d=4l+1$ and PHS breaking as $d=4l+3$. When that symmetry is broken in odd dimension, the topological invariant becomes $\bZ_2$. Therefore, if the original strong index in $\bZ$ is odd, after symmetry breaking, one of the Dirac cones is still topological protected. Here, we are interested in gapped surface states and leave the semimetal surface states for the future discussion. We focus on that the strong index is two. We find that all of the gapless states in the surface only can be gapped by the unique symmetry breaking term $S_{\bZ^C}\otimes\sigma_y$. The Pauli matrix $\sigma_y$, the breaking symmetry switches from PHS(TRS) to TRS(PHS) as $d=4l+3(4l+1)$. With TR operator $T=T_{\bZ^R}^\mathrm{surf}\otimes \bI$ and PH operator $C=C_{\bZ^R}^\mathrm{surf}\otimes \bI$ the gapped surface Hamiltonian is written as
\be
\ch^{\mathrm{surf}}_{\bZ_2^{R,2}\leftarrow\bZ^R}=\sum_{i=1}^{2n}k_i\Upsilon_i\otimes \bI+m S_{\bZ^R}^\mathrm{surf}\otimes\sigma_y,
\ee 
The first term is the gapless surface Hamiltonian and the size of $\Upsilon_i$s is $2^n\times 2^n$. In $2n$-dimension, the minimum dimension of a bulk Dirac Hamiltonian is $2^{n+1}\times 2^{n+1}$ by \cref{evensize}. Therefore, this Hamiltonian shows that the system has a half topological physics.

\subsubsection{$\bZ\leftarrow \bZ_2$}
In odd dimension $d=2n+1$, there are two possibilities to have non-trivial topological surface physics after breaking one symmetry --- $\bZ^C\leftarrow\bZ_2^{R,1}$ and $2\bZ^R\leftarrow \bZ_2^{R,2}$. By observing the classification table, the two cases in the same dimension break the same symmetry. That is, $d=4l+1$ corresponds to PHS breaking and $d=4l+3$ corresponds to TRS breaking. We add the unique symmetry breaking term ($S_{\bZ^R}^\mathrm{surf}$ ) into the gapless surface Hamiltonian 
\begin{align}
\ch^{\mathrm{surf}}_{\bZ^C\leftarrow\bZ_2^{R,1}}=&mS_{\bZ^R}^\mathrm{surf}+\sum_{i=1}^{2n}k_i\Upsilon_i \\
\ch^{\mathrm{surf}}_{2\bZ^R\leftarrow \bZ_2^{R,2}}=&mS_{\bZ^R}^\mathrm{surf}\otimes\bI+\sum_{i=1}^{2n}k_i\Upsilon_i\otimes \bI.
\end{align}
We note that the size of $\Upsilon_i$ is $2^n\times 2^n$. In $d=2n$ the minimum matrix dimension of bulk Dirac Hamiltonians in class $\bZ_{C}$ and $2\bZ^R$ are $2^n\times 2^n$ and $2^{n+1}\times 2^{n+1}$ respectively. The systems for these two cases have one half topological physics.  

In even dimension $d=2n+2$, the only possible way to have a non-trivial topological surface is $\bZ^C\leftarrow \bZ_2^{R,1}$. The symmetry class corresponding to $\bZ_2^{R,1}$ has TRS, PHS, and chiral symmetry. \Cref{expressionHbulk} shows that the corresponding symmetry operators for the surface Hamiltonian are 
\begin{small}
\begin{align}
T=T_{\bZ^R}^\mathrm{surf}\otimes \bI,\ C=T^\mathrm{surf}_{\bZ^R}\otimes \sigma_x,\ &\text{and}\ S=\bI\otimes\sigma_x\ \text{as } d=4l\\
T=C_{\bZ^R}^\mathrm{surf}\otimes \sigma_y,\ C=C^\mathrm{surf}_{\bZ^R}\otimes \bI,\  &\text{and}\ S=\bI\otimes\sigma_y\ \text{as } d=4l+2
\end{align}
\end{small}
\noindent We can add the unique mass term $m\dI\otimes \sigma_y$ and $m\dI\otimes\sigma_x$ respectively, which breaks TRS and PHS but preserve chiral symmetry, into the gapless surface Hamiltonian 
\begin{align}
\ch^{\mathrm{surf}}_{\bZ^C\leftarrow \bZ_2^{R,1}}&=\sum_{i=1}^{2n+1}k_i\Upsilon_i\otimes \sigma_z+m\bI\otimes \sigma_y,\ \text{as } n=2l-1  \\
\ch^{\mathrm{surf}}_{\bZ^C\leftarrow \bZ_2^{R,1}}&=\sum_{i=1}^{2n+1}k_i\Upsilon_i\otimes \sigma_z+m\bI\otimes \sigma_x,\ \text{as } n=2l
\end{align}
The size ($2^{n+1}\times 2^{n+1}$) of the surface Hamiltonian in the symmetry class of $\bZ_2^{R,1}$ is consistent with the minimum matrix dimension of the bulk Dirac Hamiltonian for $\bZ^C$ in $d=2n+1$ with a mass term. Hence, the system has one half topological physics.

\subsubsection{$\bZ_2\leftarrow \bZ_2$}
In odd dimension $d=2n+1$, there is only one case of nontrivial topological physics after one symmetry breaking --- $\bZ_2^{R,2}\leftarrow\bZ_2^{R,2}$. From the classification table, the dimension $d=4l+1$ corresponds to TRS breaking and the dimension $d=4l+3$ corresponds to PHS breaking. We are seeking this symmetry breaking term. First, the corresponding symmetries are $T_{\bZ_2^{R,2}}=T_{\bZ^R}\otimes \sigma_y$ and $C_{\bZ_2^{R,2}}=C_{\bZ^R}\otimes \bI$ as $d=4l+1$ and $T_{\bZ_2^{R,2}}=T_{\bZ^R}\otimes\bI$ and $C_{\bZ_2^{R,2}}=C_{\bZ^R}\otimes\sigma_y$ as $d=4l+3$. Also, from $\bZ\leftarrow \bZ$ case, $S_{\bZ^R}$ breaks PHS as $d=4l+1$ and TRS as $d=4l+3$. Therefore, the unique mass term $mS_{\bZ^R}\otimes\sigma_y$ exchanges TRS and PHS breaking and preserve the symmetry corresponding to $\bZ_2^{R,2}$ in $d=2n$. The surface Hamiltonian can be added by the projection of this breaking term from the bulk 
\be
\ch^{\mathrm{surf}}_{\bZ_2^{R,2}\leftarrow\bZ_2^{R,2}}=\sum_{i=1}^{2n}k_i\Upsilon_i\otimes\bI+mS_{\bZ^R}^\mathrm{surf}\otimes \sigma_y.
\ee
Similarly, the size of surface Hamiltonian $2^{n+1}\times 2^{n+1}$ fits the minimum matrix dimension of a bulk Dirac Hamiltonian of $\bZ_2^{R,2}$ in $d=2n$. Thus, the system possesses a half topological physics. 

In even dimension $d=2n+2$, the surface Hamiltonian in \cref{expressionHbulk} have the corresponding two symmetry operators: $T_{\bZ^1_R}=T_{\bZ^R}\otimes\bI$ and $C_{\bZ_2^{R,1}}=T_{\bZ^R}\otimes \sigma_x$ as $d=4l$ and $T_{\bZ^1_R}=C_{\bZ^R}\otimes\sigma_y$ and $C_{\bZ_2^{R,1}}=T_{\bZ^R}\otimes \bI$ as $d=4l+2$. Observing $\bZ_2^{R,1}\leftarrow \bZ_2^{R,1}$ in the periodic table, we find that PHS is broken as $d=4l$ and TRS is broken as $d=4l+2$. We can gap the surface Hamiltonian by adding the unique symmetry breaking mass term obeying the rules above
\be
H^{\mathrm{surf}}_{\bZ_2^{R,1}\leftarrow \bZ_2^{R,1}}=\sum_{i=1}^{2n+1}k_i\Upsilon_i\otimes \sigma_z+ \bI\otimes \sigma_k,
\ee
where $k=x$ as $d=4l$ and $k=y$ as $d=4l+2$. We note that the size of the surface Hamiltonian is $2^n\times 2^n$. Also, it is the same with the minimum matrix dimension of a bulk Dirac Hamiltonian for $\bZ_2^{R,1}$ in $d=2n+1$. This gapped Hamiltonian exhibits a half topological property.

\section{Generalization of a non-trivial surface on a {\it weak\/} topological insulator}\label{weakreduction}

	Gapped by CDW, the surface of a 3D weak topological insulator exhibits a half quantum spin Hall effect in \cref{gapWTI}. This CDW keeps the system in the same symmetry class by preserving TRS. In this section, we generalize this idea to some weak topological insulators and superconductors that possess a pair of the Dirac cones at two different locations of the surface Brillouin zone. The surface states of such systems are gapless but unstable in the presence of density wave. By observing the periodic table, there are two possible procedures to have topological surfaces: $\bZ_2^{R,2}\leftarrow \bZ_2^{R,1}$ and $\bZ_2^{R,1}\leftarrow \bZ^R$.

	A single Dirac cone $\ch^{\mathrm{surf}}_{\mathrm{Dirac}}$ is topologically protected in class $\bZ$ and $\bZ_2$ in the absence of symmetry-preserving gap-opening terms. However, for the surface of class $\bZ_2$ possessing two Dirac cones, there exists a symmetry-preserving coupling between these two cones so that the surface becomes gapped. The case of class $\bZ$ is also true if the two Dirac cones have the opposite orientations. That is, the surface Hamiltonian $H^{\mathrm{surf}}$ of the two Dirac cones can be gapped out by a off-diagonal term
\be
\ch^{\mathrm{surf}}=
\begin{pmatrix}
\ch^{\mathrm{surf}}_{1\mathrm{Dirac}} & 0 \\
0 & \ch^{\mathrm{surf}}_{2\mathrm{Dirac}}
\end{pmatrix}.
\ee	 
If these two cones are at the same lattice momentum, some symmetry preserving perturbations can easily gap out the surface. This is the reason that such system is classified as trivial. However, in our case these two cones are at the different lattice momenta. Only scattering being their lattice momentum difference can be the coupling between the two cones. In the following of this section, we will identify the explicit form of such coupling gapping $H^{\mathrm{surf}}$ for $\bZ_2^{R,2}\leftarrow \bZ_2^{R,1}$ and $\bZ_2^{R,1}\leftarrow \bZ^R$ in any spatial dimension and discuss the topological features of the gapped surface states. However, a trivial ($``0"$) symmetry class may have weak indices,\cite{weakindices} if in the lower dimension this class has a $\bZ$ topological invariant. It is possible to consider $\bZ\leftarrow 0$. Due to its complexity, we leave this part as future direction.

\subsection{$\bZ_2^{R,1}\leftarrow\bZ^R$}
In general, $\bZ^R$ weak topological insulators and superconductors may have even number of gapless cones with the opposite orientations on the surface. For simplicity, we put our focus on a pair of the gapless cones. When we write down the surface Hamiltonian in $d$ spatial dimension, the opposition orientations can expressed by the different signs of $\Upsilon_1$ without loss of generality 
\be
\ch^{\mathrm{surf}}_{\bZ_2^{R,1}\leftarrow \bZ^R}=
\begin{pmatrix}
k_1\Upsilon_1+\sum_{i=2}^{d-1}k_i\Upsilon_i & 0 \\
0 & -k_1\Upsilon_1+\sum_{i=2}^{d-1}k_i\Upsilon_i \label{HbZRtwocones}
\end{pmatrix}.
\ee
The two gapless cones are located at different momenta so the off-diagonal terms between these two gapless cones must come from density waves. In odd(even) spatial dimension, the Hamiltonian preserves TRS and(or) PHS with $T=\bI\otimes T_{\bZ^R}$ and(or) $C=\bI\otimes C_{\bZ^R}$. In \cref{real}, by comparing with the minimum dimension of surface Hamiltonian $\ch^{\mathrm{surf}}_{\bZ_2^{R,1}\leftarrow \bZ^R}$ and bulk Hamiltonian $\ch_{\bZ_2^{R,1}}$ in the same dimension, there exists a unique symmetry-preserving gap-opening term $m\sigma_y\otimes \Upsilon_1$. The reason is that if there more one gap opening terms, $H_{\bZ_2^{R,1}}$ is classified to $``0"$, which contradicts $\bZ_2$ invariant. With $m\sigma_y\otimes \Upsilon_1$, $H^{\mathrm{surf}}_{\bZ_2^{R,1}\leftarrow \bZ^R}$ is equivalent to $H_{\bZ_2^{R,1}}$. We can make a similar argument with \cref{strongreduction}; the surface system exhibits a half topological physics feature.

\subsection{$\bZ_2^{R,2}\leftarrow \bZ_2^{R,1}$}
For class $\bZ_2^{R,1}$, the surface Hamiltonians of a gapless cone in odd and even spatial dimensions have different forms of the expression $H^{\mathrm{surf}}_{\bZ^R}$ from \cref{expressionHbulk}; the surface physics in odd and even dimensions should be considered separately. In odd dimension ($d=2n+1$), the surface Hamiltonian of a weak topological insulator or superconductor possesses two gapless cones
\be
\ch_{\bZ_2^{R,2}\leftarrow \bZ_2^{R,1}}^{\mathrm{surf}}=
\begin{pmatrix}
\sum_{i=1}^{2n}k_i\Upsilon_i & 0 \\
0 & \sum_{i=1}^{2n}k_i\Upsilon_i \label{HbZ2twoconeseven}
\end{pmatrix}.
\ee
For class $\bZ_2$, all of the surface gapless cones are in one equivalent representation. There exists a unitary transformation so that the different expressions of two arbitrary surface gapless cones can be written in the same manner in $H_{\bZ_2^{R,2}\leftarrow \bZ_2^{R,1}}^{\mathrm{surf}}$. As $d=4l+1(4l+3)$, the Hamiltonian preserves PHS(TRS) with the operator $C=\bI\otimes C_{\bZ^R}(T=\bI\otimes T_{\bZ^R})$. Knowing the symmetry property of $S_{\bZ^C}$ from \cref{sameclass}, we find the unique symmetry preserving gap term $\sigma_y\otimes S_{\bZ^C}$. The reason is that $\sigma_y$ exchanges the preserving and breaking symmetry of $S_{\bZ_C}$. $H_{\bZ_2^{R,2}\leftarrow \bZ_2^{R,1}}^{\mathrm{surf}}$ have the same dimension(size) of $H_{\bZ_2^{R,2}}$ in $d-1$ spatial dimension. Therefore, these two Hamiltonians are equivalent so this surface system has a half topological physics feature. 

	In even spatial dimension ($d=2n+2$), based on the \cref{expressionHbulk}, the surface Hamiltonian of two gapless cones can be written in this form after the proper unitary transformation 
\be
H_{\bZ_2^{R,2}\leftarrow \bZ_2^{R,1}}^{\mathrm{surf}}=
\begin{pmatrix}
\sum_{i=1}^{2n+1}k_i\Upsilon_i\otimes\sigma_z & 0 \\
0 & \sum_{i=1}^{2n+1}k_i\Upsilon_i\otimes\sigma_z \label{HbZ2twoconesodd}
\end{pmatrix}
\ee
The system for class $\bZ_2^{R,1}$ preserves TRS and PHS; \cref{expressionHbulk} shows that as $d=4l(4l+2)$, the symmetry operators are $T=T_{\bZ^R}\otimes\bI$ and $C=T_{\bZ^R}\otimes \sigma_x$($T=C_{\bZ^R}\otimes\sigma_y$ and $C=C_{\bZ^R}\otimes \bI$). We can find the unique gap term respecting the symmetries 
\be
H_{\bZ_2^{R,2}\leftarrow \bZ_2^{R,1}}^{\mathrm{surf}}=
\begin{pmatrix}
0 & -i\bI\otimes \sigma_k \\
i\bI\otimes \sigma_k & 0
\end{pmatrix},
\ee
where $k=y(x)$ as $d=4l(4l+2)$. Likewise, in even dimension, $H_{\bZ_2^{R,2}\leftarrow \bZ_2^{R,1}}^{\mathrm{surf}}$ is equivalent to $H_{\bZ_2^{R,2}}$ in 2n+1 dimension. This system also has a topological physics feature.

\section{conclusion} \label{conclusion}

	In this paper we have studied bulk and surface Dirac Hamiltonians in Altland-Zirnbauer symmetry classes by Clifford algebra. Although the ten-fold classification has been well studied, we have provided the matrix dimensions and expressions of minimal Hamiltonians for each symmetry class and any spatial dimensions in \cref{tableminisize,expressionHbulk}. That is, a toy model can be quickly established for any topological insulators and superconductors by consulting this paper.   
	
}

	To determine non-triviality of the gapped surface states of strong or weak topological insulators and superconductors, we can directly compare with the minimum dimension of Dirac bulk Hamiltonians as shown in \cref{tableminisize} and gapless surface Hamiltonians, which are half the size of the corresponding bulk Hamiltonians in one spatial dimension higher than the surface dimension. When in a symmetry class the minimum dimension of the surface Hamiltonian is equal to or less than the minimum dimension of the bulk Hamiltonian in some symmetry classes, this surface Hamiltonian gapped by symmetry breaking or some density wave to those symmetry classes may have a nontrivial topological phases. The physical realization might be surfaces of 3D topological insulators, which has been discussed in this paper. Another possible realization is surfaces of 3D time reversal topological superconductors in class DIII. A singe Dirac cone surface mode in a strong TR topological superconductor can be gapped by a time reversal breaking term and falls into class D. The gapped surface system becomes non-trivial 2D topological superconductors described by a half Chern number. Two Dirac cone surface modes that are located at different lattice momenta with opposite orientations in a weak TR topological superconductor can be gapped by a symmetry preserving density wave. The surface physics can be realized as a \emph{half} non-trivial 2D TR superconductor.

\acknowledgments
The author thank M.~Franz, T.~Hughes, S.~Ryu, M.~Stone and J.~Teo   
for useful discussions. 
The support of the Max-Planck-UBC Centre for Quantum Materials is gratefully acknowledged. This work was mostly done in the University of Illinois at Urbana-Champaign.

\appendix
\section{The proof of the isomorphism} \label{isomorphism}
	
	We start at irreducible matrix representation of the real Clifford algebra. Consider a set of square real matrices $\{J_1,J_2,...,J_p,\tilde{J}_1,\tilde{J}_2,...,\tilde{J}_q\}$, where $J_i$ and $\tilde{J}_j$ are the generators of the Clifford algebra $Cl_{p,q}$ and obey the anticommutation relations in \cref{Cliffordrelation}. \Cref{dqp} shows the minimal dimension of $J_i$ and $\tilde{J}_j$ real matrices. For example, $p=2,\ p=1$: we can find the smallest matrices obeying those anticommutation relations are  given by
	\begin{equation}
	J_1=\sigma_x\otimes i\sigma_y,\ J_2=\sigma_z\otimes i\sigma_y, \ \tilde{J}_1=\sigma_y\otimes \sigma_y
	\end{equation}
The minimal dimension ($4\times 4$) of the matrices is consistent with the number in \cref{dqp}. On one hand, the subscript $2$ for some $Cl_{p,q}$ indicates two inequivalent representations. For example, $p=3,\ q=0$
\begin{equation}
J_1=\sigma_x\otimes i\sigma_y,\ J_2=\sigma_z\otimes i\sigma_y,\ J_3=\pm i\sigma_y \otimes \bI
\end{equation}
The signs indicate two inequivalent representations. That is, these two sets of $J_i$ cannot be orthogonal transformed from one to the other. On the other hand, without the subscript $2$ any two generator sets of $Cl_{p,q}$ can be orthogonal transformed from one to the other. 

\begin{table} 
\begin{center}
\begin{tabular}{|c||c||c|c|c|c|c|}
\hline
Class & $Cl_{p,q}$ & $q=0$ & 1 & 2 & 3 & 4  \\
\hline
\hline
D & $p=1$ & 2 & 2 & $2_2$ & 4 & 8  \\
\hline
DIII & 2 & 4 & 4 & 4 &  $4_2$ & 8  \\	
\hline
AII & 3 & $4_2$ & 8 & 8 & 8 & $8_2$  \\ 
\hline
CII & 4 & 8 & $8_2$ & 16 & 16 & 16  \\
\hline
C & 5 & 8 & 16 & $16_2$ & 32 &  32 \\
\hline
CI & 6 & 8 & 16 & 32 & $32_2$ &  64\\
\hline
AI & 7 & $8_2$ & 16 &  32& 64 & $64_2$ \\
\hline
BDI & 8 & 16 & $16_2$ & 32 & 64 & 128 \\
\hline
\end{tabular}
\end{center}
\caption{This table shows the minimum dimensions of $J_i$'s and $\tilde{J}_j$'s. The integer $p$ is the number of $J_i$'s and $q$ is the number of the $\tilde{J}_j$'s. The table easily can be extended for large $q$ and $p$ by exploiting the periodicity identity: $d_{p+8,q}=d_{p,q+8}=16d_{p,q}$.}
\label{dqp}
\end{table}

To prove those isomorphisms, we separate the proof to two following subsections. First, we consider $Cl^g_{p,q}$, a set of the quadratic forms in the real vector space, to construct the Hamiltonian in a complex vector space by using the complex structure from $J_1$. We find that there exists a mapping $f$ so that $f:Cl^g_{p,q}\rightarrow G_\#$. Secondly, we start from Dirac Hamiltonians for each individual symmetry class. Since every complex vector space has a real structure, by using this property, the inverse mapping $f^{-1}$ is found. This is our strategy to prove the isomorphisms.

\subsection{$f:Cl^g_{p,q}\rightarrow G_\#$}
	The goal of this section is to find a mapping from $Cl^g_{p,q}$ on a real vector space to $H_{\text{symmetry class}}$ on a complex vector space. To have a complex space from a twice-as-big real space, we need to introduce a complex structure. A complex structure on a real vector space is a linear map $J$ such that $J^2=-1$. In our case, we pick up $J_1$ in $Cl^g_{p,q}$ as the role of $J$ here. Moreover, $J_1=J$ acts like $``i"$. That is, for all $x, y\in \mathbb{R}$, $xv+yJ_1(v)$ in the real space $V_R$ corresponds to $xv+yiv$ in the complex space $V_C$. Also, there is an important theorem to describe linear and antilinear mappings on a complex vector space 
	
\begin{Theorem} 
A real linear transformation $A:V_R\rightarrow V_R$ is a complex linear(antilinear) transformation of the corresponding complex vector space $V_C$ if and only if $A$ commutes(anticommute) with $J$. 
\end{Theorem}	 
We will construct Hamiltonians in a real vector space from the elements of $Cl^g_{p,q}$. The Hamiltonians, which is a complex linear transformation, must commute with $J_1$. Also, a antilinear transformation in the complex vector space is a matrix anticommuting with $J_1$. Time reversal operator and particle-hole operator are antilinear and a product of an odd number of the $J_{i\neq 1}$'s and $\tilde{J}_j$'s anticommutes with $J_1$. Hence, this product is a candidate of these symmetry operators. Likewise, a product of an even number of the $J_{i\neq 1}$'s and $\tilde{J}_j$'s commutes $J_1$ so the Hamiltonian can be constructed from this product. In the following, we find a mapping for each individual symmetry class.  
	
	$f:Cl^g_{3+D,d}\rightarrow G_{\text{AII}}$: In the Hamiltonian each piece, which is an element in $G_{\text{AII}}$, is constructed from the elements of $Cl^g_{3+D,d}$ as 
\begin{equation}
H({\bf k})_{\text{AII}}=mJ_1J_{2}J_3+\sum_{j=1}^Dm_jJ_1J_{2}J_{3+j}+\sum_{i=1}^dk_iJ_{2}\tilde{J}_i. \label{HAII}
\end{equation}	
We construct a mapping so that $J_1J_2J_{3+j}\rightarrow \tilde{\gamma}_j$ for $0\leq j \leq D$ and $J_2\tilde{J}_i\rightarrow \tilde{\tilde{\gamma}}_i$ for $1\leq i \leq d$. The mapping is proper, since the matrices of $J_1J_2J_{i+3}$ and $J_2\tilde{J}_j$ have the anticommutation relation in \cref{gammaanti}. To determine which symmetry class this Hamiltonian belongs to, we try to find a product of an odd number of the $J_{i\neq 1}$'s and $\tilde{J}_j$'s obeying the symmetry equations \cref{time,PH}. Only $J_2$ is proper time reversal operator $T$ satisfying \cref{time}. Because $J_2^2=-\dI$, this Hamiltonian in a complex vector space belongs to class AII. 

$f:Cl^g_{4+D,d}\rightarrow G_{\text{CII}}$: similarly, the Hamiltonian is the same form in \cref{HAII}. We found that the Hamiltonian preserves two symmetries with time reversal operator $J_2\rightarrow T$ and particle-hole operator $J_{4+D}\rightarrow C$. Hence, the corresponding symmetry class is class CII. 
	
$f:Cl^g_{2+d,1+D}\rightarrow G_{\text{C}}$: we change the construction of the Dirac Hamiltonian from $\tilde{J}_i$ and $J_i$
\begin{equation}
H({\bf k})_{\text{C}}=mJ_2\tilde{J}_1+\sum_{j=1}^Dm_jJ_{2}\tilde{J}_{1+j}+\sum_{i=1}^dk_iJ_1J_{2}J_{i+2} \label{HC}
\end{equation}
Consider the mapping from the real quadratic forms to the complex quadratic forms: $J_2\tilde{J}_{j+1}\rightarrow \tilde{\gamma}_j$ for $0\leq j \leq D$ and $J_1J_2\tilde{J}_{i+2}\rightarrow \tilde{\tilde{\gamma}}_i$ for $1\leq i \leq d$. The anticommutation relations in \cref{gammaanti} are still preserved for $J_2\tilde{J}_{j+1}$ and $J_1J_2\tilde{J}_{i+2}$. By looking at all possible candidates of the symmetry operators, the Hamiltonian only preserves PHS as $J_2\rightarrow C$ PH operator. The Hamiltonian belongs to class C.
 
$f:Cl^g_{2+d,2+D}\rightarrow G_{\text{CI}}$: likewise, we keep the Hamiltonian in the same form in \cref{HC}. The system has another symmetry --- TRS with $\tilde{J}_{2+D}\rightarrow T$. Therefore, the corresponding symmetry class is class CI.

$f:Cl^g_{1+d,2+D}\rightarrow G_{\text{AI}}$: the Hamiltonian is in the form of 
\be
H({\bf k})_{\text{C}}=mJ_1\tilde{J}_1\tilde{J}_2+\sum_{j=1}^Dm_jJ_1\tilde{J}_{1}\tilde{J}_{2+j}+\sum_{i=1}^dk_i\tilde{J}_1J_{i+1}.
\ee
The mapping is $J_1\tilde{J}_1\tilde{J}_{2+j}\rightarrow \tilde{\gamma}_j$ and $\tilde{J}_1J_{1+i}\rightarrow \gamma_i$. The only possible symmetry operator is the TR operator ($\tilde{J}_{1}\rightarrow T$). Hence, the Hamiltonian belongs to class AI. 

$f:Cl^g_{1+d,3+D}\rightarrow G_{\text{BDI}}$: The Hamiltonian is the same form with $H({\bf k})_{\text{C}}$. The extra symmetry operator is the PH operator ($\tilde{J}_{1}\rightarrow C$). This is class BDI. 

$f:Cl^g_{2+D,1+d}\rightarrow G_{\text{D}}$: The form of the Hamiltonian is 
\be
H({\bf k})_{\text{D}}=m\tilde{J}_1J_2+\sum_{j=1}^Dm_j\tilde{J}_1J_{2+j}+\sum_{i=1}^dk_iJ_1\tilde{J}_1\tilde{J}_{1+i},
\ee
with PH operator $\tilde{J}_1\rightarrow C$. Likewise, the mapping to the complex vector space is $\tilde{J}_1\tilde{J}_{2+j}\rightarrow \tilde{\gamma}_j$ and $J_1\tilde{J}_1J_{1+i}\rightarrow \gamma_i$. Hence, the symmetry class is class CI. 

$f:Cl^g_{3+D,1+d}\rightarrow G_{\text{DIII}}$: The Hamiltonian is unchanged with the extra TRS $\tilde{J}_{3+D}\rightarrow T$. Therefore, it corresponds to class DIII.

\subsection{$f^{-1}:G_{\text{symmetry class}}\rightarrow Cl^g_{p,q}$}\label{toreal}

\begin{widetext}

\begin{table} 
\begin{center}
\begin{tabular} {|c | c | c | c|}
\hline
Class & Symmetry operators & Mapping & $Cl_{p,q}$ \\
\hline
\hline
D & $\Gamma_C^2=\bI$ & $\tilde{J}_1=\Gamma_C,\ J_{j+2}=\Gamma_C\tilde{\Gamma}_j,\ \tilde{J}_{i+1}=J_1\Gamma_C\Gamma_i$ & $Cl_{2+D,1+d}$ \\
\hline
DIII & $-\Gamma_T^2=\Gamma_C^2=\bI$ & $J_2=\Gamma_T,\ \tilde{J}_1=\Gamma_C,\ J_{j+3}=\Gamma_C\tilde{\Gamma}_j,\ \tilde{J}_{i+1}=J_1\Gamma_C\Gamma_i$ & $Cl_{3+D,1+d}$ \\
\hline
AII & $\Gamma_T^2=-\bI$ & $J_2=\Gamma_T,\ J_{j+3}=J_1\Gamma_T\tilde{\Gamma}_j,\ \tilde{J}_{i}=\Gamma_T\Gamma_i$ & $Cl_{3+D,d}$ \\
\hline
CII & $\Gamma_T^2=\Gamma_C^2=-\bI$ & $J_2=\Gamma_T,\ J_3=\Gamma_C,\ J\ J_{j+4}=J_1\Gamma_T\tilde{\Gamma}_j,\ \tilde{J}_{i}=\Gamma_T\Gamma_i$ & $Cl_{4+D,d}$ \\
\hline
C & $\Gamma_C^2=-\bI$ & $J_2=\Gamma_C,\ J_{i+2}=J_1\Gamma_C\Gamma_i,\ \tilde{J}_{j+1}=\Gamma_C\tilde{\Gamma}_j$ & $Cl_{2+d,1+D}$ \\
\hline
CI & $\Gamma_T^2=-\Gamma_C^2=\bI$ & $J_2=\Gamma_C,\ \tilde{J}_1=\Gamma_T,\ J_{i+2}=J_1\Gamma_C\Gamma_i,\ \tilde{J}_{j+2}=\Gamma_C\tilde{\Gamma}_j$ & $Cl_{2+d,2+D}$ \\
\hline
AI & $\Gamma_T^2=\bI$ & $\tilde{J}_1=\Gamma_T,\ J_{i+1}=\Gamma_T\Gamma_i,\ \tilde{J}_{j+2}=J_1\Gamma_T\tilde{\Gamma}_j$ & $Cl_{1+d,2+D}$ \\
\hline
BDI & $\Gamma_T^2=\Gamma_C^2=\bI$ & $\tilde{J}_1=\Gamma_T,\ \tilde{J}_2=\Gamma_C,\ \ J_{i+1}=\Gamma_T\Gamma_i,\ \tilde{J}_{j+3}=J_1\Gamma_T\tilde{\Gamma}_j$ & $Cl_{1+d,3+D}$ \\
\hline
\end{tabular} 
\end{center}
\caption{ Based on the anticommutation and commutation relation $\{J_1,\Gamma_{T,C}\}=0$, $[\tilde{\Gamma}_j,\Gamma_T]=0$, $\{\Gamma_i,\Gamma_T\}=0$, $\{\tilde{\Gamma}_j,\Gamma_C\}=0$, $[\Gamma_i,\Gamma_C]=0$, and $\{\Gamma_C,\Gamma_T\}=0$, as $0\leq j\leq D$ and $0<i\leq d$, we construct the mapping from any symmetry class to the generators of the Clifford algebra so that $J_k$'s and $\tilde{J}_l$'s obey the anticommutation relations in \cref{Cliffordrelation}. }
\label{HtoCl}
\end{table}

\end{widetext}

We construct the mapping from all of the elements in the Dirac Hamiltonian in form of \cref{Hmk} and the corresponding symmetry operator to the generators of the Clifford Algebra. All of these gamma matrices anticommute with each other and the square of any of them is identity. Because each complex vector space has real structure, we can easily transform from this complex vector space to a twice-big real vector space by change a complex number to a $2\times 2$ real matrix  
\begin{equation}
a+bi \rightarrow a
\begin{pmatrix}
1 & 0 \\
0 & 1
\end{pmatrix}
+b
\begin{pmatrix}
0 & 1 \\
-1 & 0
\end{pmatrix}.
\end{equation}
Since $i^2=-1$, we define 
\be
J_1=i\sigma_y\otimes \dI_{n\times n}, \label{imag}
\ee
where $n$ is the dimension of the complex vector space. The Hamiltonian in the real vector space is rewritten as 
\be
H({\bf k})=M\tilde{\Gamma}_0+\sum_{j=1}^Dm_j\tilde{\Gamma}_j+\sum_{i=1}^dk_i\Gamma_i.
\ee
Here we use the big case for the real vector space to distinguish the small case for the complex vector space. The anticommutation relation of the gamma matrices in \cref{gammaanti} can be rewritten as
\be
\{\Gamma_i,\Gamma_k\}=2\delta_{ik}\mathbb{I}_{2n\times 2n},\ \{\tilde{\Gamma}_k,\tilde{\Gamma}_j\}=2\delta_{kj}\mathbb{I}_{2n\times 2n},\ \{\Gamma_i,\tilde{\Gamma}_j\}=0
\ee
Another advantage to discuss the symmetry classes in the real vector space is that any antilinear operator can be defined in a real matrix form. Without loss of generality, we define the conjugate operator $K$ as 
\be
K=\sigma_z\otimes \dI_{n\times n},
\ee
since $K$ anticommutes with $J_1$. Hence, time reversal and particle-hole operators both are transformed to real matrix forms as $\Gamma_T$ and $\Gamma_C$ in the real vector space. Physically, there is no commutation or anticommutation relation between these two symmetry operators because they describe different degrees of freedom. Furthermore, in a complex vector space these antilinear operator do not change symmetry properties as the operators have a extra phase. Therefore, we can choose two proper antilinear operators by adding some phases so that 
\be
\{\Gamma_C,\Gamma_T\}=0.
\ee 
With the ingredients above, we can construct the mapping from a symmetry class to $Cl_{p,q}$ as follows.    

Class AII: we have some anticommutation and commutation relations $\{J_1,\Gamma_T\}=0$, $[\tilde{\Gamma}_j,\Gamma_T]=0$, and $\{\Gamma_i,\Gamma_T\}=0$. Also, $\Gamma_T^2=-\dI$. We can define $J_2=\Gamma_T$, $J_{j+3}=J_1\Gamma_T\tilde{\Gamma}_j$, $\tilde{J}_i=\Gamma_T\Gamma_i$ for $0\leq j\leq D$ and $0<i\leq d$. These $J_j$'s and $\tilde{J}_i$'s, obeying the anticommutation relation in \cref{Cliffordrelation}, construct $Cl_{3+D,d}$. 

Class CII: we still keep TRS and introduce PHS in the system. The PHS operator $\Gamma_C$ satisfies anitcommutation and commutation relations: $\{J_1,\Gamma_C\}=0$, $\{\tilde{\Gamma}_j,\Gamma_C\}=0$, and $[\Gamma_j,\Gamma_C]=0$. $\{\Gamma_T,\Gamma_C\}=0$. Also, in class CII $\Gamma_C^2=-\bI$. To satisfy \cref{Cliffordrelation} the definition of $\tilde{J_j}$ is the same with class AII; moreover, new definitions are $J_3=\Gamma_C$ and $J_{j+4}=J_1\Gamma_T\tilde{\Gamma}_j$. Those $J_i$'s and $\tilde{J}_j$'s correspond to $Cl^g_{4+D,d}$.

	For the remaining six symmetry classes, the construction of the mapping is shown in \cref{HtoCl}.

 \vskip 0.4cm

\section{Equivalent Representation} \label{equivalent}

The set $\mathcal{S}_{\bZ}$, including the elements in \cref{elements} for a Dirac Hamiltonian in any symmetry class possessing a $\bZ$ topological invariant, has two inequivalent representations. Likewise, in any symmetry class possessing a $\bZ_2$ topological invariant, the set $\mathcal{S}_{\bZ_2}$  has only one equivalent representation. More precisely, if $g_1$ and $g_2$ are in the equivalent representation, then there exists a unitary transformation $U$ so that $g_1=U^\dagger g_2 U$. In other words, any two different set of a Dirac Hamiltonians with symmetry operators in the minimal model in the same $\bZ_2$ invariant symmetry class and spatial dimension. One set can be unitary transformed to the other. For $\bZ$ it is also true, when these two sets in the equivalent representation. The proof in the following. 

      The elements in \cref{elements} can be transformed to the real vector space by applying the same transformation in \cref{toreal}
\be
\{\tilde{\Gamma}_0,\ \tilde{\Gamma}_1,\cdots,\ \tilde{\Gamma}_D,\ \Gamma_1,\ \Gamma_2,\cdots,\ \gamma_d,\ \Gamma_T,\ \text{and}\ \Gamma_C\} \label{belements}.
\ee
Now $\Gamma_T$ and $\Gamma_C$ are real matrices without the complex conjugation. We notice that $J_1$ is a surrogate of $i$ from the original complex vector space. The set above can construct all of the generators of the corresponding $CL_{p,q}$ except for $J_1$ vice versa 
\be
\{J_2,...,J_p,\tilde{J}_1,\tilde{J}_2,...,\tilde{J}_q\} \label{Clg}
\ee 
Consider the two different sets $g_1$ and $g_2$ of \cref{elements} in the complex vector space in the same symmetry class and spatial dimension. We can find the two corresponding sets of \cref{Clg}. In \cref{tableminisize}, the real Clifford algebra corresponding to $\bZ_2$ has only one equivalent representation and $\bZ$ has two inequivalent representations. Therefore, in the equivalent representation the two sets of \cref{Clg} can be {\it orthogonal \/} transformed from one to the other by an orthogonal matrix $O(2n)$. However, these two systems share the same $J_1$, which gives the restriction the orthogonal matrix
\be 
J_1=O^TJ_1O 
\ee
To obey this restriction, when the orthogonal matrix $O(2N)$ is transformed back to the original vector space, $O(2N)$ must be a unitary matrix $U(N)$\cite{Stone:2011qo}. By performing the unitary transformation $U(N)$, $g_1=U^\dagger g_2 U$.

\bibliographystyle{apsrev4-1}
\bibliography{TOPO3}

\end{document}